\documentclass[epj]{svjour}
\usepackage{cite}
\usepackage{bm}
\usepackage{amssymb}
\usepackage[normalem]{ulem}
\usepackage[dvips]{graphicx}
\usepackage{scalefnt}
\begin{document}
\title{Quark matter under strong magnetic fields\thanks{Special Issue on ``Exotic Matter in Neutron Stars".}}
\author{D\'ebora Peres Menezes\inst{1} \and Luiz La\'ercio Lopes\inst{1,2}
}                     

\institute{Depto de F\'{\i}sica - CFM - Universidade Federal de Santa
Catarina -  Florian\'opolis - SC  - CEP 88.040 - 900 - Brazil \and 
Centro Federal de Educa\c{c}\~ao Tecnol\'ogica de Minas Gerais, Campus VIII - Varginha - MG - CEP 37.022-056 - Brazil}
\date{Received: date / Revised version: date}
%
\abstract{We revisit three of the mathematical formalisms used to
  describe magnetized quark matter in compact objects within the MIT
  and the Nambu-Jona-Lasinio models and then compare their
  results. The tree formalisms are based on 1) isotropic equations of
  state, 2) anisotropic equations of state with different parallel and
  perpendicular pressures and 3) the assumption of a chaotic field
  approximation that  results in a truly  isotropic equation of state.
We have seen that the magnetization obtained with both models
is very different: while the MIT model produces well-behaved curves
that are always positive for large magnetic fields, the NJL model yields a
magnetization with lots of spikes and negative values. This fact has
strong consequences on the results based on the existence of anisotropic 
equations of state. We have also seen that, while the isotropic
formalism results in maximum stellar masses that increase considerably
when the magnetic fields increase, maximum masses obtained with the
chaotic field approximation never vary more than 5.5$\%$. 
The effect of the magnetic field on the radii is opposed in the MIT
and NJL models: with  both formalisms, isotropic and chaotic field
approximation, for a fixed
mass, the radii increase with the increase of the magnetic field in
the MIT bag model and decrease in the NJL, the radii of quark stars
described by the NJL model being smaller than the ones described by
the MIT model.
} 
\maketitle
\section{Introduction}
\label{intro}

The influence of magnetic fields ($B$) in the formation, constitution and
evolution of stars is a topic of intense investigation and discussion.
Some astronomers believe that magnetism plays an essential role in
the star formation, second only to gravitation \cite{zirker}. As the star
burns its fuel, depending on its mass, it may end as a white dwarf or
a neutron star and in both cases, magnetic fields can be quite
intense. Neutron stars are very dense compact objects with magnetic
fields generally of the order of $10^{12}$ G and they are mainly
detected as pulsars, powered by rotation energy, or as accreting
X-ray binaries, powered by gravitational energy. However, since 1979,
other classes of neutron stars have been observed as either soft
gamma-ray repeaters or as anomalous X-ray pulsars with no binary
companion. These objects bear magnetic fields of the order of $10^{14}
- 10^{16}$ G and were named magnetars \cite{magnetars}.

If one observes the  QCD phase diagram, neutron stars lie in a region
of low temperatures and high baryonic chemical potentials $\mu$ (or baryonic
densities, if one prefers). If these objects are affected by the
presence of strong magnetic fields, other regions of the QCD phase
diagram should be affected as well \cite{qcd+b},\cite{CEP1}.

The region of high temperatures and low
chemical potentials is the relevant regime for the study of heavy ion
collisions \cite{kharzeev} and in these experiments, the 
field intensity probably decreases very rapidly, lasting for about 1-2
fm/c only,  but a possible signature of the strong magnetic fields has been 
investigated \cite{tuchin},\cite{mclerran}, \cite{event}, \cite{paoli}, \cite{photon_anisotropy}.
Moreover, this region of high-$T$/low-$\mu$,  has already  been
exploited using Lattice (LQCD) simulations.
The QCD phase diagram presents a first order transition at 
low-$T$/high-$\mu$ region and a crossover region at 
high-$T$/low-$\mu$ and these two regions may be connected by a
critical end point \cite{CEP1},\cite{CEP2}.  LQCD results point out, in
accordance with most model predictions, that the crossover observed
at $B=0$ does not disappear when $B \ne 0$ \cite{earlylattice},\cite{preussker},\cite{lattice1},\cite{lattice2}. 
Nevertheless the region of low-$T$/high-$\mu$ remains unavailable to
LQCD assessments and hence, the physical properties related to this
region can only be investigated with the help of (effective) models. 

Other important aspects related to magnetized quark matter are the 
richness of its internal substructures and the inverse magnetic
catalysis (IMC).  Although the results are model dependent, the internal structure of 
the QCD phase diagram has been shown to
be very rich \cite{structure}, presenting lots of intermediate
phases, due to the small jumps in the quark dressed masses, which are
related to the filling of the Landau levels. 
Concerning the IMC, LQCD calculations predict an inverse catalysis,
with  the pseudo-critical temperature decreasing with $B$ \cite{lattice1,lattice2}, 
while most effective models predict an increase of the pseudo-critical
temperature with $B$ \cite{catalysis-ours}, \cite{ricardo}. It is a common belief that
this deficiency in effective models is due to the fact that they do not account for back reaction
effects (the indirect interaction of gluons with the magnetic field)
\cite{teoriaIMC}.

One important point that should not be disregarded refers to the
contribution of the electromagnetic interaction to the pressure(s) and
energy density, a term proportional do $B^2$, where $B$ is the
magnetic field strength. Astrophysicists advocate that the magnetic
field in the surface of magnetars is of the order of $10^{15}$G. However, 
according to the Virial theorem one could expect fields three or four
orders of magnitude stronger in their interior. To cope with this
difference in strength, an  {\it ad~hoc} exponential density-dependent
magnetic field was proposed in ref.~\cite{Pal} and adopted in 
many other works~\cite{Mao},\cite{Rabhi}, \cite{prc2}, \cite{rudiney},
\cite{Ryu}, \cite{Rabhi2}, \cite{Mallick}, \cite{luizlaercio},\cite{Dex},\cite{njlv},
\cite{Mallick2},\cite{Ro}.  Another similar prescription for an {\it energy density
dependent} magnetic field was proposed in Ref. \cite{chaotic}. 
The second  prescription seems more natural because the quantity used in the
TOV equations \cite{tov} to calculate the macroscopic quantities is
the energy density and not the number density. Moreover, in
\cite{chaotic}, it is shown that this choice reduces the number of
free parameters and it is given by:

\begin{equation}
B (\epsilon) =   B_0 \bigg ({\frac{\epsilon_M}{\epsilon_0}} \bigg
)^{\gamma} + B^{surf}, 
\label{s8}
\end{equation}
where  $B_0$ is the fixed value of the magnetic field, $\epsilon_0$  is the energy density at the center of the maximum mass neutron star
with zero magnetic field,  $\epsilon_M$ is the energy density of the matter alone, avoiding
a self-generated magnetic field, and $\gamma$ is any positive number. In the present work we
take $\gamma=3$ and thus, we reduce the number of free 
parameters from two, used in the prescription proposed in \cite{Pal} to only one. Nevertheless, 
as pointed in ref.~\cite{chaotic}, for chaotic magnetic fields, Eq.~(\ref{s8}) yields a parameter-free model,
while for the standard density-dependent magnetic field, the macroscopic properties are strongly dependent of these
two non-observables parameters.
Another point worth mentioning is that the magnetic field is no longer fixed for all neutron star configurations.
Each equation of state (EOS) produces a different value for $\epsilon_0$ that enters in  Eq.~(\ref{s8}).
{ In this sense, $\epsilon_0$ is also another parameter, but it is
  not arbitrary as $\gamma$ and the two parameters that can be arbitrarily
changed in the parametrization proposed in ref.~\cite{Pal}. It comes directly from the
model that defines the EOS, as the baryonic density, pressure and
energy density that enter as input to the TOV \cite{tov}. }

One should bear in mind that these two {\it ans\"atze}, violate Maxwell
equations { because the divergent of $B$ is no longer zero}.  
To minimize this problem, in the present work only the
contribution to the term proportional to $B^2$ will be taken as
density dependent. The EOS obtained from magnetized matter will always
carry a fixed $B$ value. 
{ With this choice, if we apply the models to describe stellar
  matter,  we guarantee thermodynamical consistency in the
  EOS and still force matter at the surface of the star to be subject to a
  magnetic field that is not too strong. It is also important to
  stress that if a density dependent magnetic field is used through out the calculations,
numerical results are almost coincident with the ones we present next
but, besides violating Maxwell equations, thermodynamical consistency would also require an
extra term, normally disregarded in the literature.}
This point will be specifically emphasized
when the equations are displayed in the next sections.

In the present work we study quark matter possibly present in strange
(quark) stars. According to the Bodmer-Witten conjecture
\cite{Itoh},\cite{bodmer},\cite{witten}, the interior of a neutron-like star does
not consist primarily of hadrons, but rather of quark matter composed of deconfined up, down 
and strange quarks, plus the leptons necessary to ensure charge neutrality  and $\beta$-equilibrium. 
When a model is chosen, it is important to verify if it
satisfies the necessary conditions that ensure that quark matter is
the true ground state: two-flavor quark matter must be unstable
(i.e., at zero temperature its energy per baryon has to be larger than
930 MeV, the iron binding energy) and the three-flavor quark matter
must be stable (i.e., its energy per baryon must be lower than 930
MeV, also at $T=0$).  Not all quark matter models satisfy these
conditions and the NJL model only provides absolute stable matter when
it is magnetized \cite{stabilityB}. As already said, according to the Virial theorem
and assuming uniform field and mass density, the maximum magnetic
field allowed in a gravitationally bound star is smaller than  $ 10^{19}$ G.
 However, a quark star is self-bound by the nuclear interaction and
a simple estimative of the maximum allowed magnetic field
gives fields of the order of  $10^{20}$ G \cite{efrain, noronha2007}.  We use these values
as a guide to our study. 

It is worth mentioning that 
calculations that take into account the effects of the anomalous
magnetic moments (AMM) of the quarks forsee that the maximum possible
magnetic field strength is 8.6 $\times 10^{17}$ G due to the fact that
the $u$ quark polarization would become complex for higher values
\cite{aurora1}. On the other hand, works on hadronic stellar matter show that the
influence of the (nucleonic and hyperonic) AMM on the EOS is
minor \cite{rudiney},\cite{rhabi1}, if the magnetic fields are of the
order of $B=10^{18}$ G. However, if $B>10^{18}$G, the inclusion of the
anomalous magnetic moment contribution stiffens the stellar matter EOS, and may
originate a total spin polarization of neutrons \cite{broderick00}. 
In the present work, we do not consider the coupling between the field and the quark
anomalous magnetic moment, but we think this problem should be tackled
in a future work. We come back to this point in the conclusions of the
paper.

Despite the fact that usually a mean field approximation is necessary
when one uses effective models, restricting their scope and the
interpretation of their results, they can be very helpful if one wants
to understand the physics not assessed by LQCD. Thus, different
models have been recently used to describe quark matter subject to
magnetic field, the most common ones being the MIT bag model
\cite{mit} and the Nambu-Jona-Lasinio model \cite{njl}. In the present
work we revisit both of them within three different formalisms 
employed in the literature to describe stellar matter. 

The first formalism considers that quark matter is iso\-tropic
\cite{blandford} and it was used in the seminal works on magnetars
described either by a quark matter \cite{chakra96} or 
by a hadronic matter equation of state \cite{lattimer_2000}  
and in subsequent calculations \cite{rhabi1},\cite{rhabi2}. 
The first calculations involving quark matter described by the NJL
model followed the same line of understanding\cite{prc2},\cite{rudiney},
\cite{luizlaercio}, \cite{prc1}.

The second formalism is based on the fact that 
the EOS  cannot be  truly isotropic, once 
the components of the energy-momentum  tensor are not equal,
giving different contributions for the parallel and perpendicular
pressure \cite{efrain}, \cite{aurora1}, \cite{sedrakian}, 
\cite{jorge}, \cite{aurora2}. { Moreover, under strong magnetic
  fields, the $O(3)$ rotational symmetry breaks, and this is another
  argument that supports the existence of pressure anisotropy.}
Once the EOS is obtained, one observes at what magnetic
field the pressures start to deviate from each other and this is the
limit usually taken as input to the Tolman-Oppenheimer-Volkoff
equations (TOV)  \cite{tov}. As pointed in ref.~\cite{sedrakian},
this limit is around $3.1 \times 10^{18}$G. In practice, the EOS used to compute the
stellar macroscopic quantities is isotropic \cite{aurora1},
\cite{jorge},\cite{veronica}. However, this approach means that one
recognizes that the system is anisotropic but ends up ignoring this fact. 
At least two recent works face this problem: in one of them
\cite{Mallick2}, the authors treated the anisotopic pressures
as a perturbation in a way similar to the Hartle-Thorn method,
generally used for slowly rotating neutron stars. In the other one
\cite{aurora2}, an axisymmetric geometry is assumed and the Einstein
equations are solved with the adoption of a cylindrical symmetric
metric. Older works also tackle this subject \cite{Bocquet}, \cite{Cardall}.
In the present work we do not intend to reproduce these more
sophisticated treatments, but we will investigate the effects of strong magnetic
fields on the anisotopic pressures. 

Finally, the underlying assumption of the third formalism is that in
the presence of anisotropies, the concept of pressure has to be taken
with care \cite{Misner,Zel}. Based on the concepts discussed in these two books, 
a small-scale chaotic field is used and the stress
tensor is modified accordingly, so that the resulting EOS is a truly isotropic one
\cite{chaotic}. 

Next, a comparison of the results obtained with the three formalisms is shown and discussed.
Although parts of this material are already available in the
literature, a correct comparison can only be made if the same choice
for the magnetic field is used. In the present work we use the energy
density dependent magnetic field given in Eq. (\ref{s8}) to compute
our results with the three formalisms. We would like to emphasize that
our aim in the present work is not to justify, criticize or choose any of the
formalisms just mentioned. We restrict ourselves to the analysis and
comparison of the results. 

The present work is organized as follows: in Section \ref{sec:2}, we
present the EOS obtained for the MIT bag model in the presence of a
magnetic field in one preferential direction with the three possible
formalisms just discussed: an isotropic EOS, an anisotropic EOS with
different parallel and perpendicular pressures and another isotropic EOS
resulting from the introduction of a chaotic field. The main results
are then shown and discussed. In Section \ref{sec:3}, the same steps
are taken for the NJL model. In Section \ref{sec:4}, some of the results
obtained with both models are compared.
 Finally, in the last Section, the final conclusions are drawn.

\section{The MIT Bag model}
\label{sec:2}

The EOS for magnetized quark matter described by the MIT bag model has
already been extensively studied \cite{efrain}, \cite{aurora1}, \cite{sedrakian},
\cite{jorge}, \cite{aurora2}. 
We next show only the Lagrangian density and the resulting EOS.
We depart from the following Lagrangian density:
\begin{equation}
{\cal L} = {\cal L}_{f}+{\cal L}_{l} - \frac {1}{4}F_{\mu
\nu}F^{\mu \nu} \,,
\label{lagran}
\end{equation}
which contains a quark sector, ${\cal L}_f$, a leptonic sector,
  ${\cal L}_l$, and the electromagnetic contribution 
$F_{\mu \nu }=\partial_{\mu }A_{\nu }-\partial _{\nu }A_{\mu }$.  We use a 
  static and constant magnetic field parallel to the $z$ direction and hence, 
we choose the gauge $A_\mu=\delta_{\mu 2} x_1 B$ and the energy levels in the $x$ and $y$
directions are quantized. 

The leptonic sector is described by
\begin{equation}
\mathcal{L}_l=\bar \psi_l\left[\gamma_\mu\left(i\partial^{\mu} - q_l A^{\mu}
\right) -m_l\right]\psi_l \,\,,
\label{lage}
\end{equation}
where $l=e,\mu$. As we restrict ourselves to the $T=0$ case, the star has no
more trapped neutrinos, which have escaped and carried a huge amount
of energy while the star cooled down. 

The thermodynamical potential for the three flavor
quark sector, $\Omega_f$, can be written as 
\begin{equation}
\Omega_f = -P_f = {\cal E}_f - \sum_f \mu_f n_f,
\label{thermo}
\end{equation} 
 where $P_f$ represents the pressure, ${\cal E}_f$ the energy density, 
$\mu_f$ the chemical potential, and $n_f$ the quark number
density.  A similar expression can be written
for the leptonic sector. 

\subsection{MIT - Isotropic EOS}

In order to obtain the complete EOS, the 
pressure, and baryonic density were calculated in 
\cite{stabilityB}, \cite{efrain}, \cite{sedrakian}  and are given by
$$
P_f =\sum_{k=0}^{k_{f,max}} \alpha_k\frac {|q_f| B N_c }{4 \pi^2}  \left [
  \mu_f \sqrt{\mu_f^2 - s_f(k,B)^2}  \right.
$$
\begin{equation}
\left.
- s_f(k,B)^2 \ln \left ( \frac { \mu_f +\sqrt{\mu_f^2 -
s_f(k,B)^2}} {s_f(k,B)} \right ) \right ] -\mathcal{B} \mbox{,}  
\label{pbag}
\end{equation}
where $N_c$ is the number of colors, $q_f$ is the electric charge of
each quark, $B$ is the magnetic field
strength and the quark masses are $m_{u,d}=5$ MeV, $m_s$=120 MeV,
$\mathcal{B}$ is the bag constant fixed as 148 MeV$^{1/4}$,
$\alpha_0=1,\,\alpha_{k>0}=2$,
\begin{equation}
s_f(k,B) = \sqrt {m_f^2 + 2 |q_f| B k}
\end{equation} 
and
\begin{equation}
n=\sum_i \frac{n_{f}}{3},
\end{equation}
with
\begin{equation} 
n_f=\sum_{k=0}^{k_{f,max}} \alpha_k \frac{|q_f| B N_c }{2 \pi^2} k_{F,f} \mbox{,}  
\label{rhobag}
\end{equation}
where $k_{F,f}=\sqrt{\mu_f^2 - s_f(k,B)^2}$.
At $T=0$, the  upper Landau level (or the nearest integer)  is defined by
\begin{equation}
k_{f, max} = \frac {\mu_f^2 -m_f^2}{2 |q_f|B}= \frac{k_{F,f}^2}{2|q_f|B},
\label{landaulevels}
\end{equation}
and the energy density can be easily obtained from
Eqs. (\ref{thermo}),(\ref{pbag}) and (\ref{rhobag}).

In the description of compact stars, both charge neutrality and
chemical equilibrium conditions have to be imposed  
\cite{Glen00}. The first condition can be written for quarks and leptons as
\begin{equation}
2n _{u}=n _{d}+n _{s}+3\left( n _{e}+n _{\mu }\right) \mbox{,}
\label{neut}
\end{equation}
and the second condition can be written as
\begin{equation}
\mu _{s}=\mu _{d}=\mu _{u}+\mu _{e}\mbox{, \ \ }\mu _{e}=\mu _{\mu } \mbox{,}
\label{qch}
\end{equation}
where the lepton densities can be calculated through
Eq.~(\ref{rhobag}),  with appropriate substitutions for the masses and electric charge.
The lepton masses are $m_e=0.511$ MeV and $m_\mu=105.66$ MeV. For the
electron and muon pressure ($P_l$) and energy density ($\epsilon_l$) 
we use Eqs. (\ref{pbag}) and the leptonic analogue of
Eq.(\ref{thermo}),  respectively, all with $\mathcal{B}=0$.
In all cases, $N_c=1$ for the leptons.

The final expressions for the total pressure and energy density are
given by

\begin{equation}
P_{iso}=P_f + P_l + B(\epsilon)^2/2,
\label{pressmitiso} 
\end{equation}
\begin{equation}
\epsilon_{iso}=\epsilon_f + \epsilon_l + B(\epsilon)^2/2 .
\label{enermitiso}
\end{equation}

\subsection{MIT - Anisotropic EOS}

The magnetization of the system is given by
\begin{equation}
{\cal M} = {dP}/{dB},
\label{mag}
\end{equation}
and for the quark sector of the MIT bag model, it reads:
$$
{\cal M}_f =\sum_{k=0}^{k_{f,max}} \alpha_k\frac {|q_f| N_c }{4 \pi^2}  \left [
  \mu_f \sqrt{\mu_f^2 - s_f(k,B)^2}  \right.
$$
\begin{equation}
\left.
- (m_f^2 + 4 |q_f| B k)
 \ln \left ( \frac { \mu_f +\sqrt{\mu_f^2 -
s_f(k,B)^2}} {s_f(k,B)} \right ) \right ] \mbox{.}  
\label{mitmag}
\end{equation}

In an anisotropic system, the parallel and the perpendicular components of the pressure can be
written in terms of the magnetization, as \cite{Mallick},
\cite{efrain}, \cite{sedrakian}, \cite{veronica}, \cite{aurora2}:

\begin{equation}
P_{\parallel}= P_f + P_l - B(\epsilon)^2/2, \quad
P_{\perp}= P_f + P_l -{\cal M}B + B(\epsilon)^2/2 \mbox{.}  
\label{pressmitaniso}
\end{equation}

For a magnetic field in the $z$ direction, the
stress tensor has the form: diag$(B^2/2,B^2/2,-B^2/2)$ and this
explains the difference in sign appearing in the parallel and
perpendicular pressures. The energy density is the same as in the
isotropic system, given by Eq. (\ref{enermitiso}).

The magnetization for the lepton sector can be read off
Eq.(\ref{mitmag}) with the appropriate substitutions specified below.

\subsection{MIT - Chaotic field}
 
According to \cite{Misner}, as the components of the stress tensor
are not equal (they differ in sign), the quantity $\pm B^2/2$ cannot be simply
added to the pressure terms. As already stated above, for a magnetic
field in the $z$ direction, the stress tensor is given by
diag$(B^2/2,B^2/2,-B^2/2)$. One way of circumventing this problem is 
 proposed in \cite{Zel}, where it is shown that, for a small scale
 chaotic field, the concept of isotropic pressure is recovered
 because the stress tensor is diag$(B^2/6,B^2/6,B^2/6)$, { the
   $O(3)$ rotational symmetry remains valid}  and
 hence, the pressure becomes:
\begin{equation}
 P =  \frac{1}{3} < T_i^i > = \frac{1}{3} \bigg ( \frac{B^2}{6} + \frac{B^2}{6}  + \frac{B^2}{6}  \bigg ) =  \frac{B^2}{6} . \label{s6}
\end{equation}
Note that  Eq.~(\ref{s6})  
represents the true thermodynamic pressure, since the components of the
stress tensor are equal~\cite{Zel}, in opposition to the two first
formalisms.
We would like to point out that due to strong heat transportation,
the existing turbulence could create a chaotic field. Of course, as
the star cools down, the fields could become ordered. The possible 
mechanisms responsible for creating and maintaining a chaotic field
would have to be better investigated and are beyond the scope of the
present work. 

Within this formalism, the expressions for the total pressure and energy density are
given by:

\begin{equation}
P_{chao}=P_f + P_l + B(\epsilon)^2/6,
\label{pressmitchao} 
\end{equation}
\begin{equation}
\epsilon_{chao}=\epsilon_f + \epsilon_l + B(\epsilon)^2/2 .
\label{enermitchao}
\end{equation}

\subsection{Results MIT} 

We next show our results for the  EOS obtained from the three
formalisms just discussed. We have used
$\epsilon_0 = 6.93$ fm$^{-3}$ in Eq. (\ref{s8}) because this value is the
central energy density of the maximum mass obtained with the MIT model
for non-magnetized matter with our choice of values for the quark masses
and the Bag constant.
Before we proceed, it is important to stress that in Eqs. (\ref{pressmitiso}),
 (\ref{pressmitaniso}) and (\ref{pressmitchao}), the
magnetic field is taken as constant in $P_f$ and $P_l$ and
Eq.(\ref{s8}) is used in the last terms only. The same holds for the
energy density terms.

\begin{figure}
\centering
\includegraphics[width=0.34\textwidth, angle =270 ]{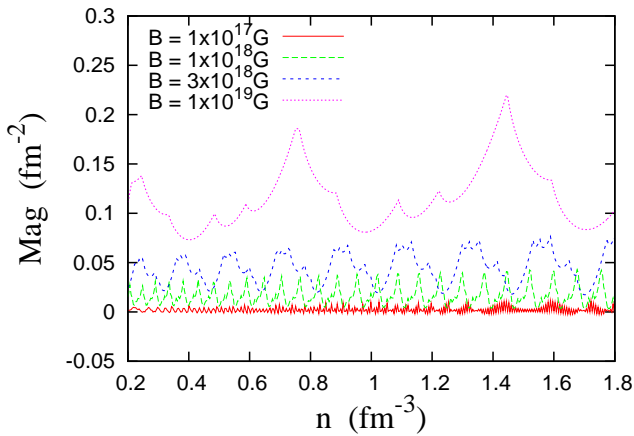}
\caption{Magnetization as a function of the number density  for the
  MIT bag model.}
\label{figmitmag}     
\end{figure}

We first analyze how the magnetization varies with density for
different values of the magnetic field in Fig. \ref{figmitmag}. It is
clear that the number of van alphen oscillations are larger for lower
values of the magnetic field as expected, due to the larger number of
filled Landau levels. As the magnetic field increases, the
magnetization of the system also increases and hence, a stronger
effect is felt on the anisotropy of the system, as can be seen in Eq. (\ref{pressmitaniso}).

\begin{figure}
\centering
\includegraphics[width=0.34\textwidth, angle =270 ]{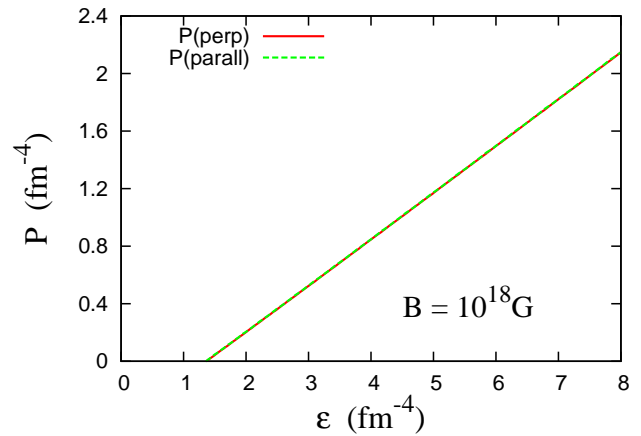}
\caption{MIT bag model EOS obtained for $B=10^{18}$ G without the term proportional to $B^2$}
\label{figmitb18}     
\end{figure}

\begin{figure}
\centering
\includegraphics[width=0.34\textwidth, angle =270 ]{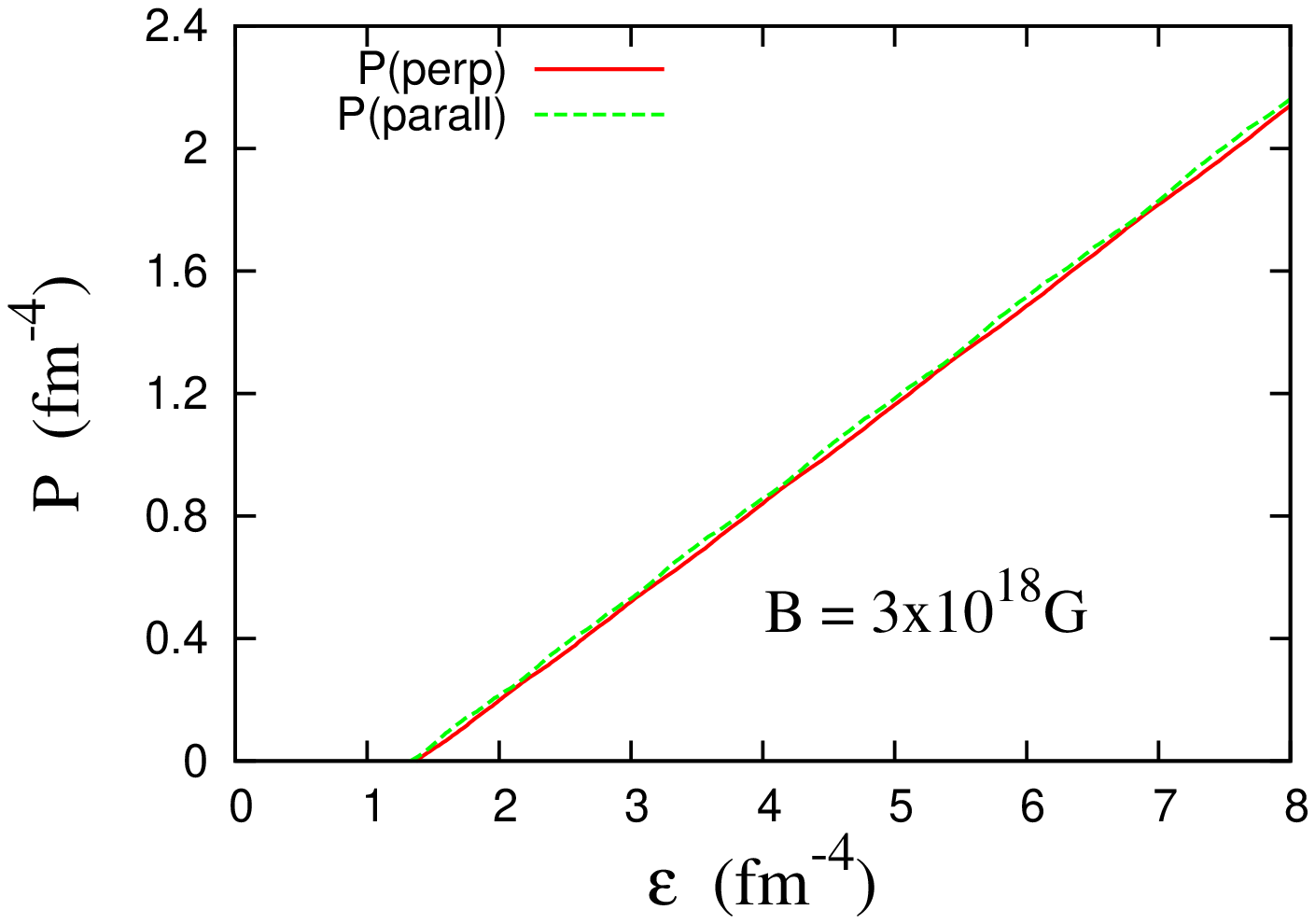}
\caption{MIT bag model EOS obtained for $B=3 \times 10^{18}$ G without the term proportional to $B^2$}
\label{figmit3b18}     
\end{figure}

\begin{figure}
\centering
\includegraphics[width=0.34\textwidth, angle =270 ]{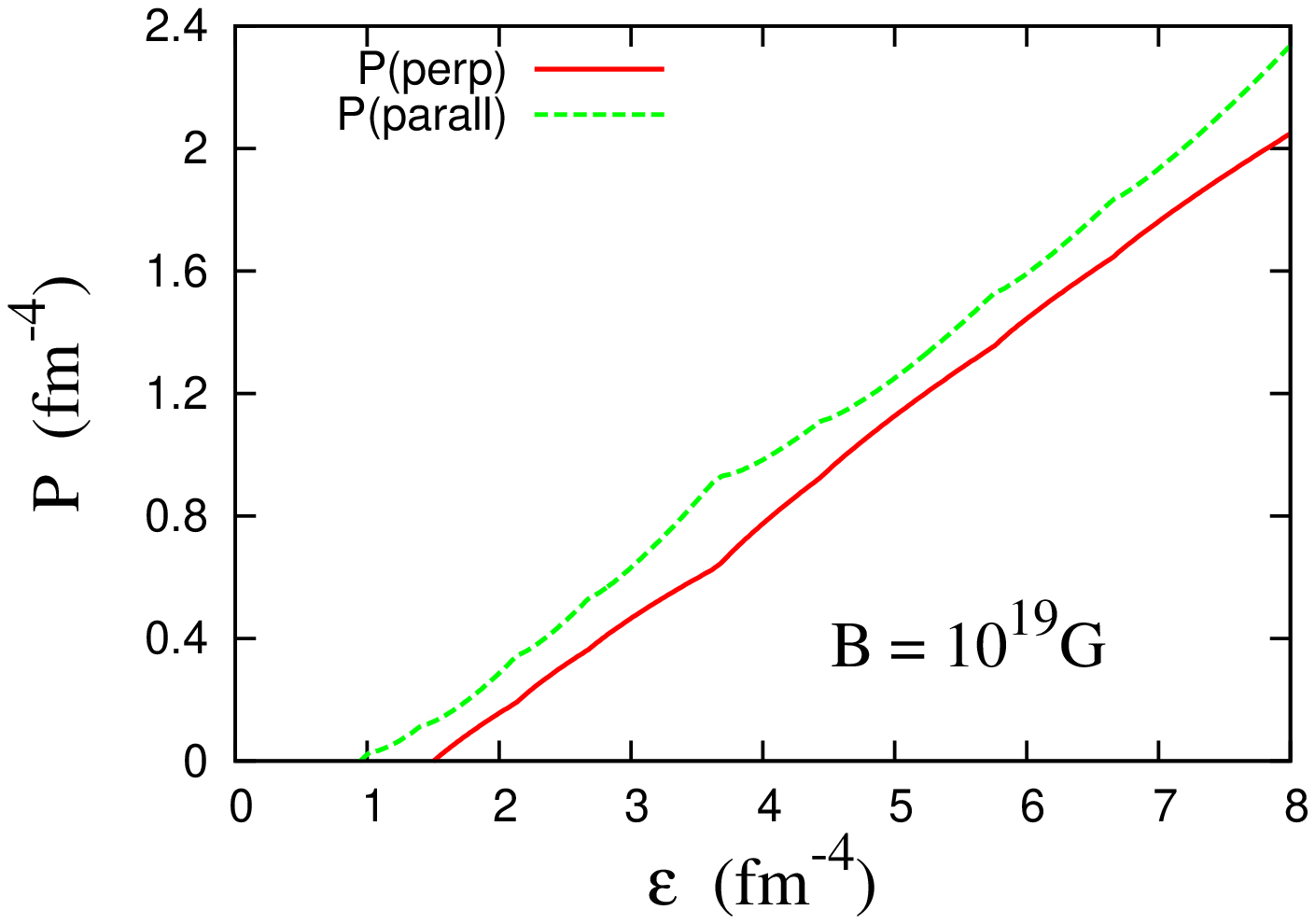}
\caption{MIT bag model EOS obtained for $B=10^{19}$ G without the term proportional to $B^2$}
\label{figmitb19}   
\end{figure}

\begin{figure}
\centering
\includegraphics[width=0.34\textwidth, angle =270 ]{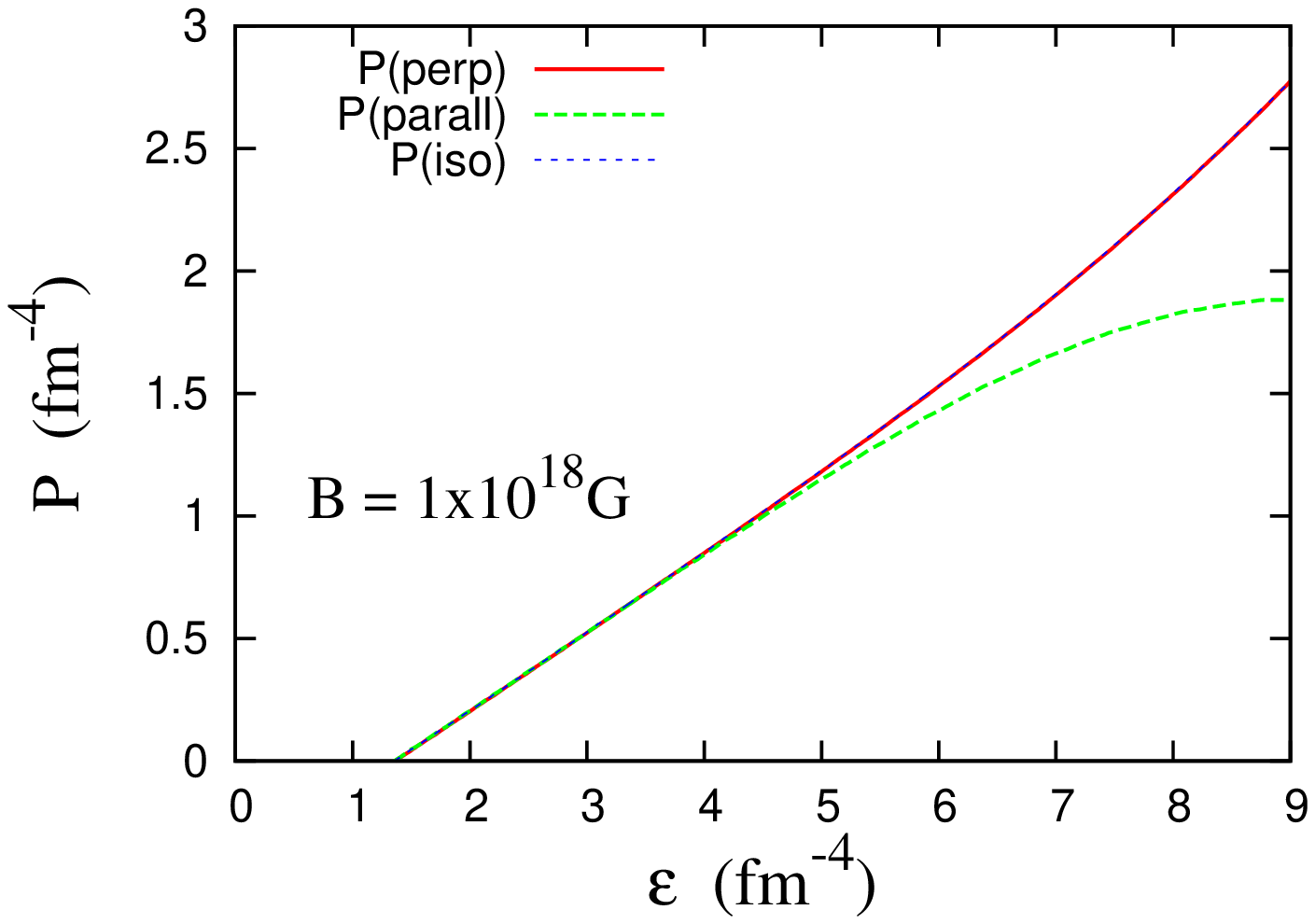}
\caption{MIT bag model EOS obtained for $B=10^{18}$ G with the term proportional to $B^2$}
\label{figmitb18a}      
\end{figure}

\begin{figure}
\centering
\includegraphics[width=0.34\textwidth, angle =270 ]{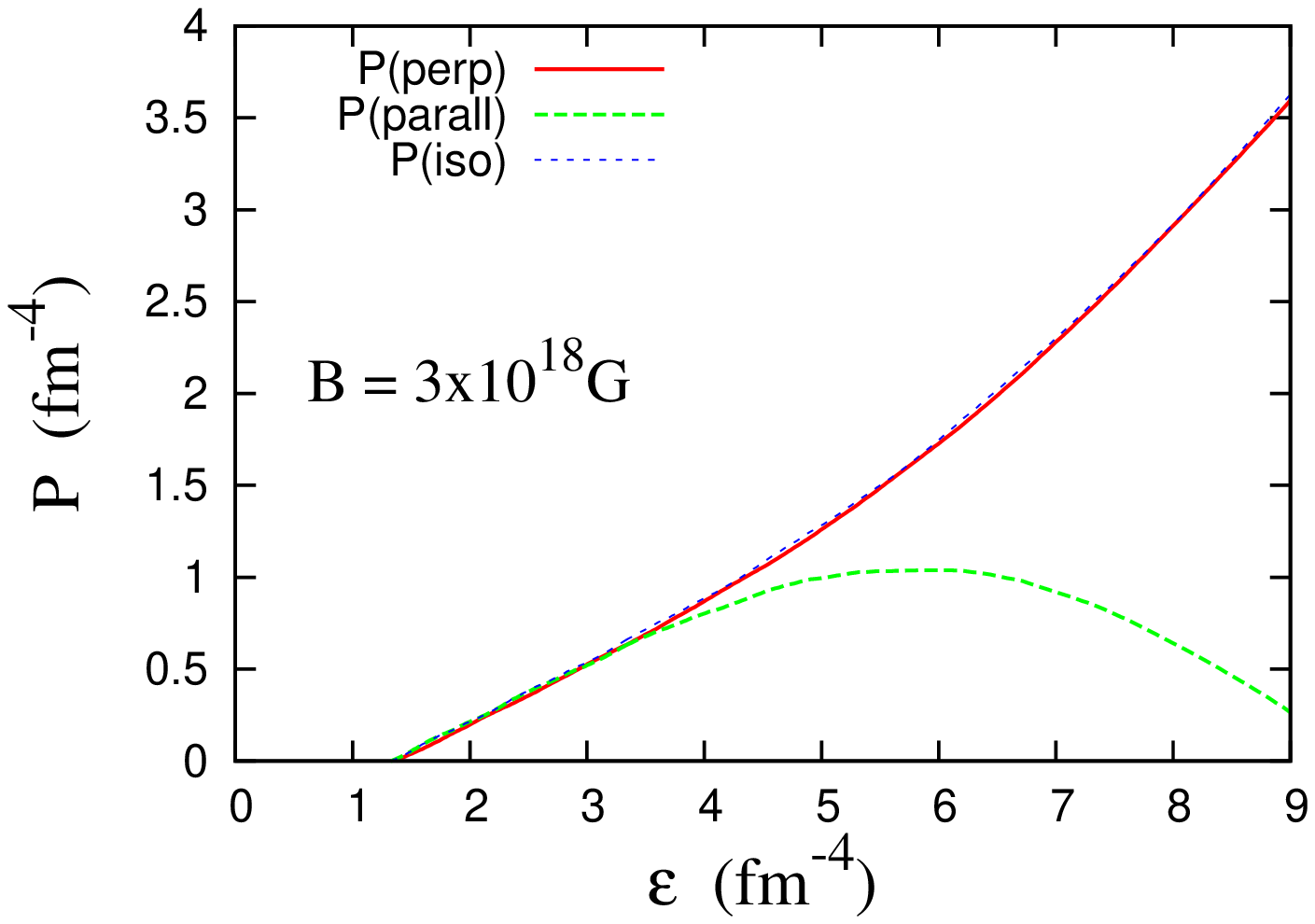}
\caption{MIT bag model EOS obtained for $B=3 \times 10^{18}$ G with the term proportional to $B^2$}
\label{figmit3b18a}      
\end{figure}

\begin{figure}
\centering
\includegraphics[width=0.34\textwidth, angle =270 ]{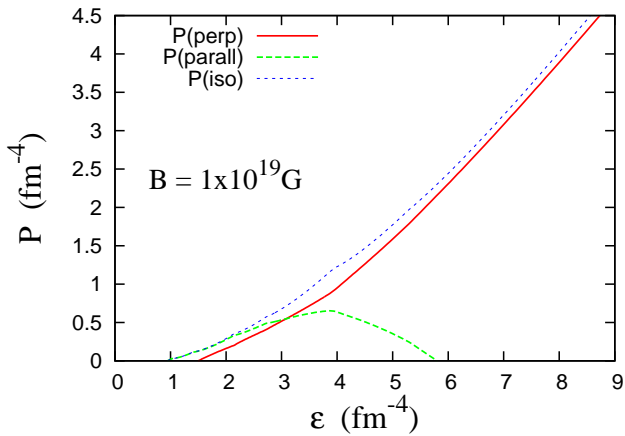}
\caption{MIT bag model EOS obtained for $B=10^{19}$ G with the term proportional to $B^2$}
\label{figmitb19a}     
\end{figure}

For $B=10^{17}$ G, the magnetization is negligible and no visible
difference can be seen when the perpendicular and parallel pressures
are plotted. Moreover, the contribution coming from the term
proportional do $B^2$ is also too small and, either it is included or
not, the plots are coincident.  For these reasons, we just plot
results for magnetic fields larger than $10^{18}$ G. In
Figs. (\ref{figmitb18}) and (\ref{figmitb18a}), we plot the parallel
and perpendicular pressures for $B=10^{18}$ G respectively without and
with the contribution proportional to the $B^2$ term. In Figs. 
(\ref{figmit3b18}) and (\ref{figmit3b18a}), 
(\ref{figmitb19}) and (\ref{figmitb19a}), the same is displayed for a
larger magnetic fields, equal to $B=3 \times 10^{18}$ G and 
$B=10^{19}$ G. By analyzing these
figures, it is evident that the contribution of the $B^2$ term has
quite drastic effects on the point where the pressures start to split
and hence, as this term is important in stellar matter EOS, it is
indeed difficult to justify the use of the TOV equations for magnetic
fields larger than $10^{18}$ G. One can also see, from Figs.  
(\ref{figmitb18a}), (\ref{figmit3b18a}) and
 (\ref{figmitb19a}) that the isotropic pressure
does not deviate much from the perpendicular pressure, the only
difference being due to the magnetization of the system, which is
still relatively small for these fields. The parallel pressure goes to
zero at energy densities typical to neutron star core if $B=10^{19}$
G, but for lower magnetic fields, the decrease starts at much higher
densities. 

\begin{figure}
\centering
\includegraphics[width=0.34\textwidth, angle =270 ]{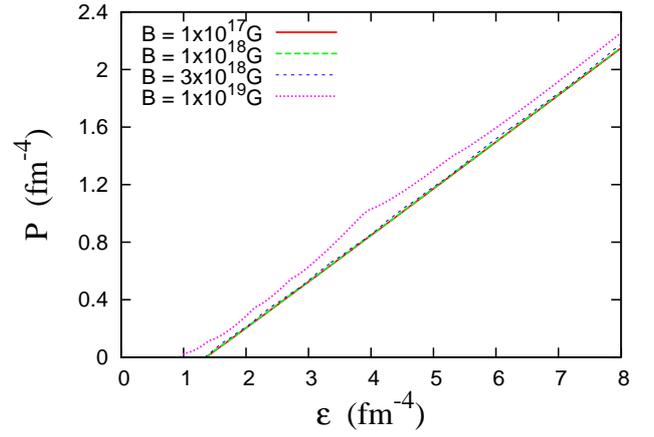}
\caption{MIT bag model EOS for different values of the chaotic magnetic field.}
\label{figmitchao}     
\end{figure}

In Fig. (\ref{figmitchao}), the EOS obtained with the assumption of
chaotic fields are shown. They deviate very little from non-magnetized
matter, with a consequent small variation in the stellar properties
that are computed next.

\begin{figure}
\centering
\includegraphics[width=0.34\textwidth, angle =270 ]{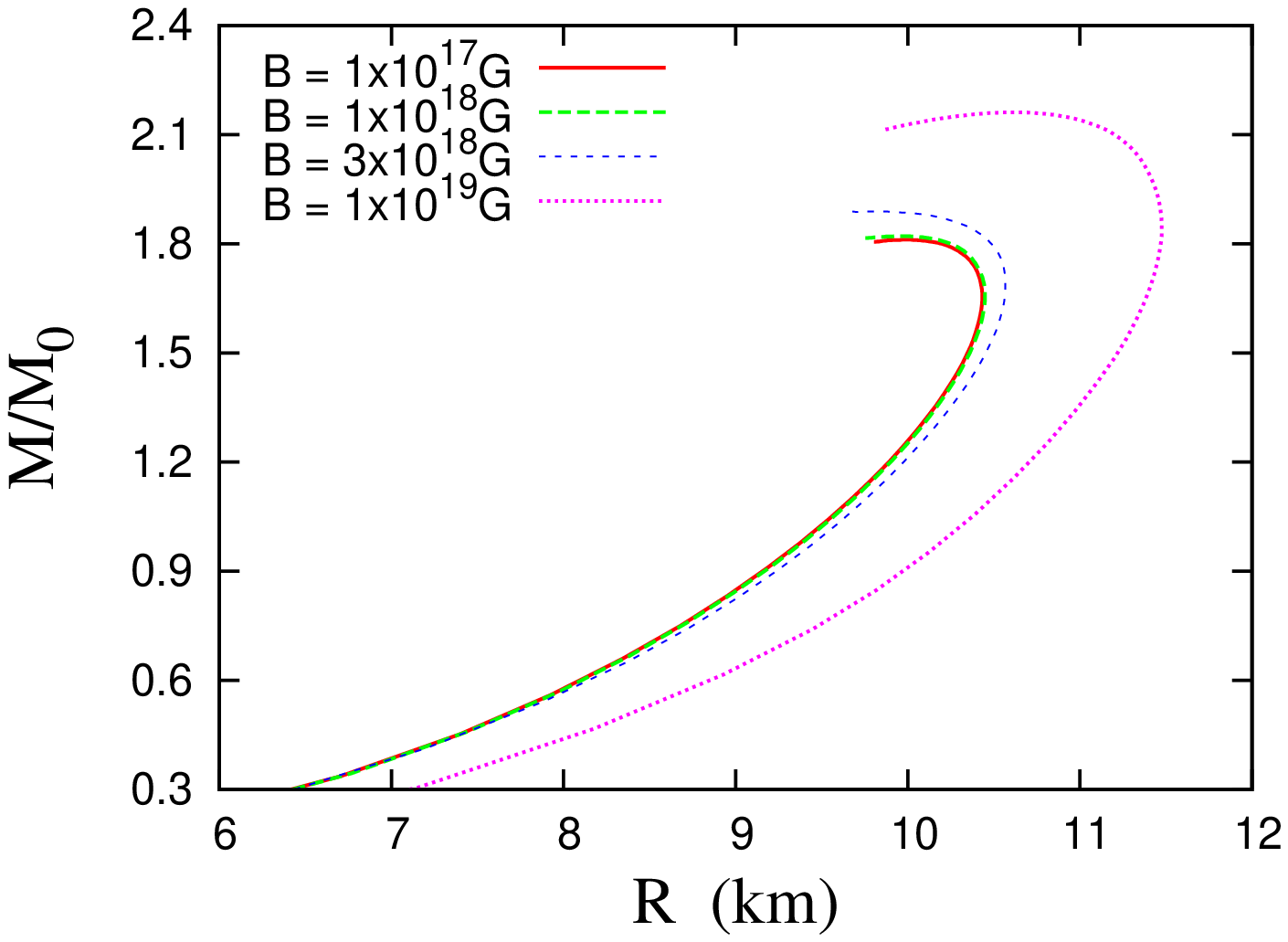}
\caption{MIT bag model: Mass-radius relation within isotropic pressure for different values of magnetic field.}
\label{tovmitiso}     
\end{figure}

\begin{figure}
\centering
\includegraphics[width=0.34\textwidth, angle =270 ]{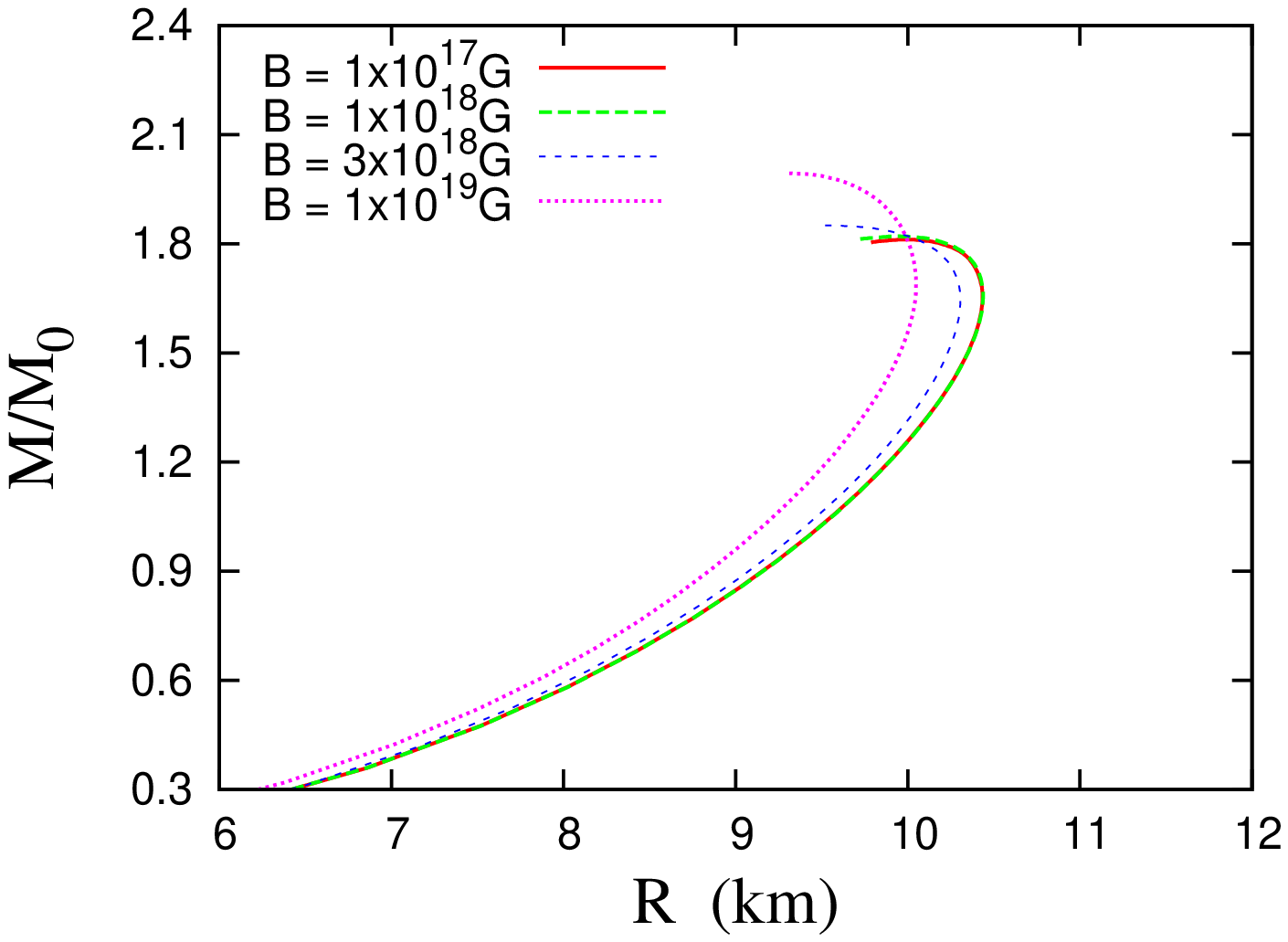}
\caption{MIT bag model: Mass-radius relation within anisotropic pressure for different values of magnetic field.}
\label{tovmitani}     
\end{figure}

\begin{figure}
\centering
\includegraphics[width=0.34\textwidth, angle =270 ]{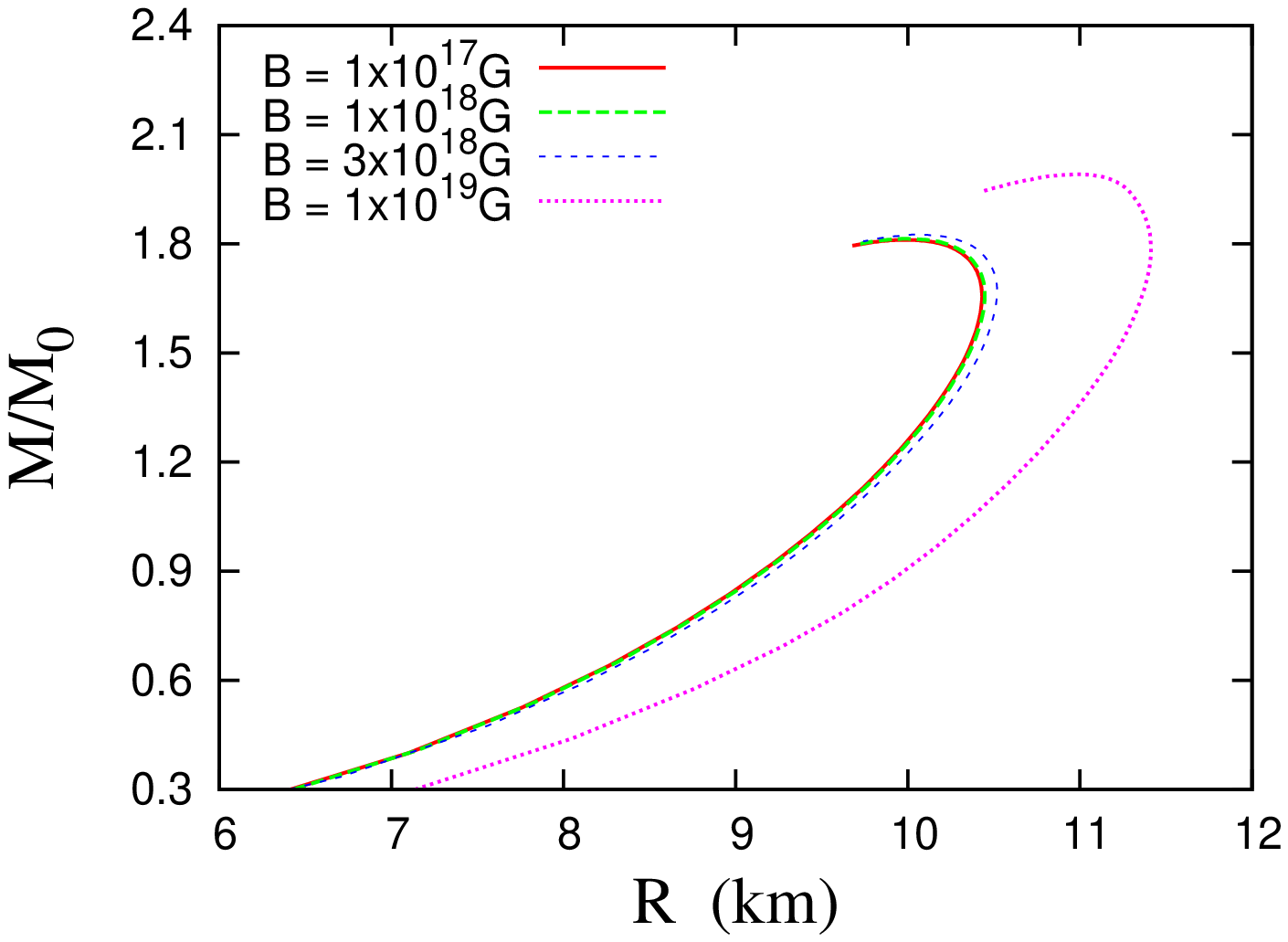}
\caption{MIT bag model:Mass-radius relation within chaotic magnetic field approximation for different values of magnetic field.}
\label{tovmitcmf}     
\end{figure}

\begin{table}
\caption{Properties of the maximum mass quark star}
\label{tabmit}       
\begin{tabular}{llllll}
\hline\noalign{\smallskip}
Model & $B$ & $M_{max}$ & $Mb_{max}$ & R & ${\cal E}_c$ \\
\noalign{\smallskip}\hline\noalign{\smallskip}
& (G) & ($M_\odot$) & ($M_\odot$) & (km) & (fm$^{-4}$) \\
\noalign{\smallskip}\hline
MIT$_{\rm isotropic}$ & $10^{17}$ & 1.81  & 2.32  & 9.99 & 6.77 \\
\noalign{\smallskip}\hline
MIT$_{\rm anisotropic}$ & $10^{17}$ & 1.81  & 2.32  & 9.99 & 6.83 \\
\noalign{\smallskip}\hline
MIT$_{\rm chaotic}$ & $10^{17}$ & 1.81  & 2.32  & 9.99 & 6.80 \\
\noalign{\smallskip}\hline
\noalign{\smallskip}\hline
MIT$_{\rm isotropic}$ & $10^{18}$ & 1.82  & 2.33  & 9.97 & 6.77 \\
\noalign{\smallskip}\hline
MIT$_{\rm anisotropic}$ & $10^{18}$ & 1.82  & 2.33  & 9.95 & 6.85 \\
\noalign{\smallskip}\hline
MIT$_{\rm chaotic}$ & $10^{18}$ & 1.82  & 2.32  & 10.03 & 6.63 \\
\noalign{\smallskip}\hline
\noalign{\smallskip}\hline
MIT$_{\rm isotropic}$ & $3. 10^{18}$ & 1.89  & 2.42  & 9.78 & 7.52 \\
\noalign{\smallskip}\hline
MIT$_{\rm anisotropic}$ & $3. 10^{18}$ & 1.85  & 2.36  & 9.58 & 7.63 \\
\noalign{\smallskip}\hline
MIT$_{\rm chaotic}$ & $3.10^{18}$ & 1.83  & 2.33  & 10.08 & 6.63 \\
\noalign{\smallskip}\hline
\noalign{\smallskip}\hline
MIT$_{\rm isotropic}$ & $10^{19}$ & 2.16  & 2.90  & 10.68 & 6.33 \\
\noalign{\smallskip}\hline
MIT$_{\rm anisotropic}$ & $10^{19}$ & 1.99  & 2.49  & 9.31 & 8.39 \\
\noalign{\smallskip}\hline
MIT$_{\rm chaotic}$ & $10^{19}$ & 1.99  & 2.66  & 10.97 & 5.87 \\
\noalign{\smallskip}\hline
\end{tabular}
\vspace*{0.8cm}  
\end{table}

We now use the EOS obtained from the three different formalisms we
have discussed to compute the macroscopic properties  with the
  help of the TOV equations \cite{tov}. In Figs. \ref{tovmitiso},
\ref{tovmitani}, \ref{tovmitcmf}  we display the mass-radius relation
for  isotropic, anisotropic and chaotic magnetic field approximation
respectively, and in Table \ref{tabmit} we display the properties
for the maximum mass quark star. Had we chosen a smaller value for
the strange quark mass, we would have obtained larger maximum masses, as
can be seen, for instance, from Fig. 5b in Ref. \cite{james}. It means
that a 2 $M_\odot$ star could be attained with another choice of
parameters, but our general conclusions remain the same. 
{ The same statement is valid with respect to the choice of the bag
  constant  $\mathcal{B}$ fixed as 148 MeV$^{1/4}$, a value that 
satisfies the Bodmer-Witten conjecture once the stability window is
investigated and gives a maximum stellar mass not too low, as 
can be seen also from Fig. 5b in Ref. \cite{james}.}

The maximum
mass only shows a real increase as compared with the one obtained from
non-magnetized matter (which is identical to the results obtained for
$B=10^{17}$ G) for $B > 3 \times 10^{18}$ G for an isotropic
EOS and for $B \simeq 10^{19}$ G if the chaotic field or the 
anisotropic pressure is used. This is
easily understood because the EOS for chaotic field approximation is
softer than the one for the isotropic EOS, as one sees by comparing
Eqs. (\ref{pressmitiso}) and (\ref{pressmitchao}). In solving the
TOV equations,  we have used the perpendicular pressure (and not the
parallel one) given in Eq.(\ref{pressmitaniso}) because the parallel
pressure goes to zero at certain energy densities, as discussed above.
Hence, for the magnetic fields of interest, the maximum mass would be
too low. This effect was also shown in \cite{aurora2}, where the
authors have computed the TOV equations with both pressures. Of
course, it is very difficult to give a reasonable physical
interpretation for this kind of calculation that only considers one of
the existing pressures, but we  present it
here for the sake of completeness.

Looking at the results shown in Table \ref{tabmit}, one can see that
the maximum stellar mass increases by approximately 20$\%$ as compared
with the non-magnetized star (equal to $10^{17}$ G if the isotropic
EOS is used and only by  5.5$\%$ with the use of the chaotic field
formalism. { Our results come directy from the fact that the EOS
  obtained with the isotropic formalism is much harder than the one
  obtained with the chaotic approximation (it has an extra 1/3 factor in the
  pressure of the magnetic field) for strong magnetic fields.}
 The results obtained with the perpendicular pressure of the
anisotropic EOS are similar to the ones obtained with the chaotic
field approximation however, always with a smaller radii. 
Indeed we see that higher the magnetization, the more compact the
quark star is.

Another subject that we would like to discuss is related to the
recent improvements of both, theory and observations 
of neutron star radii. Based on chiral effective theory, the radii of
the canonical 1.4$M_{\odot}$ neutron star radius was 
constrained to 9.7 - 13.9 Km in \cite{hebeler}. In 
\cite{Steiner1},~\cite{Steiner2} a limit of 12 km for the 1.4$M_{\odot}$ was predicted,
while in \cite{Lim} this limit was set to 13.1 km.
 On the other hand, in \cite{webb} and \cite{Steiner3}, it was
assumed that all neutron stars have the same radii and they should lie between 7.6 and 10.4
km and 10.9 and 12.7 km respectively. We resume the properties of 1.4$M_{\odot}$ quark
stars obtained with the formalisms discussed in the present work in Table~\ref{MIT14}.
We see that for all magnetic fields, the MIT bag model fulfills all
constraints, except the ones proposed in refs.\cite{webb} and
\cite{Steiner3}, because they are mutually exclusive and cannot be
satisfied at the same time.

\begin{table}
\caption{Properties of the 1.4$M_{\odot}$ quark star}
\label{MIT14}       
\begin{tabular}{llllll}
\hline\noalign{\smallskip}
Model & $B$ & $M_{max}$ & $Mb_{max}$ & R & ${\cal E}_c$ \\
\noalign{\smallskip}\hline\noalign{\smallskip}
& (G) & ($M_\odot$) & ($M_\odot$) & (km) & (fm$^{-4}$) \\
\noalign{\smallskip}\hline
MIT$_{\rm isotropic}$ & $10^{17}$ & 1.40  & 1.71  & 10.23 & 2.64 \\
\noalign{\smallskip}\hline
MIT$_{\rm anisotropic}$ & $10^{17}$ & 1.40  & 1.72  & 10.24 & 2.54 \\
\noalign{\smallskip}\hline
MIT$_{\rm chaotic}$ & $10^{17}$ & 1.40  & 1.72  & 10.24 & 2.65 \\
\noalign{\smallskip}\hline
\noalign{\smallskip}\hline
MIT$_{\rm isotropic}$ & $10^{18}$ & 1.40  & 1.72  & 10.25 & 2.65 \\
\noalign{\smallskip}\hline
MIT$_{\rm anisotropic}$ & $10^{18}$ & 1.40  & 1.72  & 10.24 & 2.65 \\
\noalign{\smallskip}\hline
MIT$_{\rm chaotic}$ & $10^{18}$ & 1.40  & 1.73  & 10.25 & 2.65 \\
\noalign{\smallskip}\hline
\noalign{\smallskip}\hline
MIT$_{\rm isotropic}$ & $3. 10^{18}$ & 1.40  & 1.73  & 10.34 & 2.53 \\
\noalign{\smallskip}\hline
MIT$_{\rm anisotropic}$ & $3. 10^{18}$ & 1.40  & 1.73  & 10.13 & 2.78 \\
\noalign{\smallskip}\hline
MIT$_{\rm chaotic}$ & $3.10^{18}$ & 1.40  & 1.74  & 10.32 & 2.60 \\
\noalign{\smallskip}\hline
\noalign{\smallskip}\hline
MIT$_{\rm isotropic}$ & $10^{19}$ & 1.40  & 1.81  & 11.10 & 2.07 \\
\noalign{\smallskip}\hline
MIT$_{\rm anisotropic}$ & $10^{19}$ & 1.40  & 1.78  & 9.85 & 2.97 \\
\noalign{\smallskip}\hline
MIT$_{\rm chaotic}$ & $10^{19}$ & 1.40  & 1.81  & 11.09 & 2.09 \\
\noalign{\smallskip}\hline
\end{tabular}
\vspace*{0.8cm}  
\end{table}

\section{The NJL model}
\label{sec:3}

The EOS of magnetized matter obtained with the NJL model has already
been extensively discussed in
\cite{prc1},\cite{prc2},
\cite{njlv},\cite{ani}. The Lagrangian density
is the same as given for the MIT bag model in Eq.(\ref{lagran}) and the
leptonic sector is also the same as described in Section \ref{sec:2}.
The quark sector is described by the  su(3) version of the
Nambu--Jona-Lasinio model
\begin{equation}
{\cal L}_f = {\bar{\psi}}_f \left[\gamma_\mu\left(i\partial^{\mu}
- {\hat q}_f A^{\mu} \right)-
{\hat m}_c \right ] \psi_f ~+~ {\cal L}_{sym}~+~{\cal L}_{det}~.
\label{njl}
\end{equation}
The ${\cal L}_{sym}$ and ${\cal L}_{det}$ terms are given by:
\begin{equation}
{\cal L}_{sym}~=~ G \sum_{a=0}^8 \left [({\bar \psi}_f \lambda_ a \psi_f)^2 + ({\bar \psi}_f i\gamma_5 \lambda_a
 \psi_f)^2 \right ]  ~,
\label{lsym}
\end{equation}
\begin{equation}
{\cal L}_{det}~=~-K \left \{ {\rm det}_f \left [ {\bar \psi}_f(1+\gamma_5) \psi_f \right] + 
 {\rm det}_f \left [ {\bar \psi}_f(1-\gamma_5) \psi_f \right] \right \} ~,
\label{ldet}
\end{equation}
where $\psi_f = (u,d,s)^T$ represents a quark field with three flavors, ${\hat m}_c= {\rm diag}_f (m_u,m_d,m_s)$ with $m_u=m_d \ne m_s$ is the corresponding (current) mass matrix while ${\hat q}_f={\rm diag}(q_u,q_d,q_s)$ is the matrix that
represents the quark electric charges. $\lambda_0=\sqrt{2/3}I$, with $I$ being the unit matrix in the three flavor space, and
$0<\lambda_a\le 8$ denote the Gell-Mann matrices. The term (${\cal L}_{det}$) is the t'Hooft  six-point  interaction 
and ${\cal L}_{sym}$ is a four-point interaction in flavor space. 
The model is non renormalizable, and as a regularization scheme for  the
divergent ultraviolet integrals  we use a sharp cut-off $\Lambda$ in
three-momentum space. In the present work,  we use the HK parametrization
proposed in \cite{hatsuda} :
$\Lambda = 631.4 \, {\rm MeV}$ , $m_u= m_d=\,  5.5\, {\rm MeV}$,
$m_s=\,  135.7\, {\rm MeV}$, $G \Lambda^2=1.835$ and $K \Lambda^5=9.29$.

The conditions of charge neutrality and $\beta$-equilibrium given in
Eqs. (\ref{neut}) and (\ref{qch}) are also enforced.

\subsection{NJL - Isotropic EOS}

In the mean field approximation the pressure can be written as
\begin{equation}
P_f = \theta_u+\theta_d+\theta_s 
-2G(\phi_u^2+\phi_d^2+\phi_s^2) + 4K \phi_u \phi_d \phi_s \,\,,
\end{equation}
where the effective quark masses can be obtained self consistently  from 
\begin{equation}
 M_i=m_i - 4 G \phi_i + 2K \phi_j \phi_k, 
 \label{mas}
\end{equation}
with $(i,j,k)$ being any permutation of $(u,d,s)$,
\begin{equation}
\theta_f= \left (\theta^{vac}_f+\theta^{mag}_f + \theta^{med}_f \right )_{M_f}\,\,,
\label{pressBmu2}
\end{equation}
where the vacuum contribution reads
\begin{equation}
\theta^{vac}_{f}=- \frac{N_c }{8\pi^2} \left \{ M_f^4 \ln \left [
    \frac{(\Lambda+ \epsilon_\Lambda)}{M_f} \right ]
 - \epsilon_\Lambda \, \Lambda\left(\Lambda^2 +  \epsilon_\Lambda^2 \right ) \right \},
\end{equation}
and with $\epsilon_\Lambda=\sqrt{\Lambda^2 + M_f^2}$, 
the finite  magnetic contribution is given by
\begin{equation}
\theta^{mag}_f= \frac {N_c (|q_f| B)^2}{2 \pi^2} \left [ \zeta^{\prime}(-1,x_f) -  \frac {1}{2}( x_f^2 - x_f) \ln x_f +\frac {x_f^2}{4} \right ]\,\,,
\end{equation}
with   $x_f = M_f^2/(2 |q_f| B)$ and $\zeta^{\prime}(-1,x_f)= d
\zeta(z,x_f)/dz|_{z=-1}$ where $\zeta(z,x_f)$ is the Riemann-Hurwitz
zeta function. 
The medium contribution can be written as
$$
\theta^{med}_f=\sum_{k=0}^{k_{f,max}} \alpha_k\frac {|q_f| B N_c }{4
  \pi^2}  \left [ \mu_f \sqrt{\mu_f^2 - s_f(k,B)^2} \right.
$$
\begin{equation}
\left. - s_f(k,B)^2 \ln \left ( \frac { \mu_f +\sqrt{\mu_f^2 -
s_f(k,B)^2}} {s_f(k,B)} \right ) \right ] ,
\label{PmuBt0}
\end{equation}
where  $s_f(k,B) = \sqrt {M_f^2 + 2 |q_f| B k}$, 
and the  upper Landau level (or the nearest integer)  is defined by
\begin{equation}
k_{f, max} = \frac {\mu_f^2 -M_f^2}{2 |q_f|B} .
\label{landaulevels2}
\end{equation}

The condensates $\phi_f$ in the presence of an external magnetic field can
be written as 

\begin{equation}
\phi_f=(\phi_f^{vac}+\phi_f^{mag}+\phi_f^{med})_{M_f} ,
\end{equation}
where
\begin{eqnarray}
\phi_f^{vac} &=& -\frac{ M_f N_c }{2\pi^2} \left [
\Lambda \epsilon_\Lambda -
 {M_f^2}
\ln \left ( \frac{\Lambda+ \epsilon_\Lambda}{{M_f }} \right ) \right ]\,\,,
\end{eqnarray}

$$
\phi_f^{mag}
= -\frac{ M_f |q_f| B N_c }{2\pi^2} \times
$$
\begin{equation}
\left [ \ln \Gamma(x_f)  -\frac {1}{2} \ln (2\pi) +x_f -\frac{1}{2} \left ( 2 x_f-1 \right )\ln (x_f) \right ] \,\,,
\end{equation}
and
\begin{equation}
\phi_f^{med}=
\sum_{k=0}^{k_{f,max}} \alpha_k \frac{ M_f |q_f| B N_c }{2 \pi^2} \ln \left ( \frac { \mu_ f +\sqrt{\mu_f^2 -
s_f(k,B)^2}} {s_f(k,B)} \right ) .
\label{MmuBt0}
\end{equation}

The corresponding leptonic contributions are the same as the ones used
for the MIT bag model.
The final expressions for the pressure and energy density are given as
in Eq. (\ref{pressmitiso}) and (\ref{enermitiso}).

\subsection{NJL - Anisotropic EOS}

The parallel and the perpendicular components of the pressure can be
written in terms of the magnetization as in Eq. (\ref{mag}). The
complete calculation is shown in \cite{ani}, but the main formulae
are given next.

For the quark sector the derivatives of the pressure with respect to
the magnetic field are:

$$ {\cal M}_f =
\frac{d P_f}{d B} = \theta_u^\prime+\theta_d^\prime+\theta_s^\prime 
-4G(\phi_u \phi_u^\prime+\phi_d \phi_d^\prime+\phi_s \phi_s^\prime) 
$$
\begin{equation}
+ 4K (\phi_u^\prime \phi_d \phi_s +\phi_u \phi_d^\prime \phi_s+ \phi_u \phi_d \phi_s^\prime) \,\,,
\end{equation}
where
\begin{equation}
 \theta^\prime_f=(\theta^{\prime\, vac}_f+\theta^{\prime \, mag}_f+\theta^{\prime \, med}_f)_{M_f} \,\,\,,
\end{equation}
and
\begin{equation}
 \phi^\prime_f=(\phi^{\prime \, vac}_f+\phi^{\prime \, mag}_f+ \phi^{\prime \, med}_f)_{M_f} \,\,\,,
\end{equation}
with
$$
\theta^{\prime \, mag}_f= 2 \frac{\theta^{mag}_f}{B} -
 \frac{N_c |q_f| M_f^2}{4\pi ^2}
$$
\begin{equation}
  \left[ \ln
 \Gamma(x_f)-\frac{1}{2} ln (2 \pi) + x_f - (x_f - \frac{1}{2}) \ln(x_f) \right]\; ,
 \end{equation}

\begin{equation}
\theta^{\prime \, med}_f= \frac{\theta_f^{med}}{B}- \frac{N_c B
  |q_f|}{2\pi^2} \sum_{k=0}^{k_{max}} \alpha_k \ln \left ( \frac{
    \mu_f + \sqrt{\mu_f^2-s_f^2}}{s_f} \right ) k|q_f|\, ,
\end{equation}

$$
\phi^{\prime \, mag}_f=\frac{\phi_f^{mag}}{B} + \frac{N_c
  M_f^2}{8\pi^2 B} \times
$$
\begin{equation}
\left \{ |q_f| B +M_f^2 \left[ \psi^{(0)}(x_f) - \ln(x_f)\right] \right \}\,\,,
\end{equation}
where $\psi^0(x_f)=\frac{\Gamma^{\prime}(x_f)}{\Gamma(x_f)}$ is the
digamma function. The in medium contribution reads
$$
\phi^{\prime \, med}_f= \frac{\phi_f^{med}}{B} +
$$
\begin{equation}
-\frac{  N_c |q_f|B}{2 \pi ^2}
\sum_{k=0}^{k_{f,max}} \alpha_k   \frac{
  \mu_f M_f (k |q_f|}{s_f(k,B)^2 \sqrt{\mu_f^2-s_f(k,B)^2}}  ,
\label{phimed}
\end{equation}

and $\theta^{\prime \,vac}_f$ and $\phi^{\prime \,vac}_f$ vanish.

For the leptonic sector one easily gets 
$$ {\cal M}_l=
\frac{d P^{ med}_l}{dB}= \frac{P_l^{med}}{B}
$$
\begin{equation}
- \frac{B |q_l|}{2\pi^2} \sum_{k=0}^{k_{max}} \alpha_k \ln \left (
  \frac{ \mu_l + \sqrt{\mu_l^2-s_l^2}}{s_l} \right ) (k|q_l|) \;,
\end{equation}
which is another way of expressing Eq. (\ref{mitmag}) with the
appropriate substitutions for the leptons.

The total parallel and perpendicular pressures are given as in Eq.(\ref{pressmitaniso}).

\subsection{NJL - Chaotic field}

Again, as in the MIT model, the final expressions for the EOS are
given by Eqs.(\ref{pressmitchao}) and (\ref{enermitchao}).

\subsection{Results NJL} 

We start by showing the EOS obtained from the three
formalisms within the NJL model. We have used
$\epsilon_0 = 7.81$ fm$^{-3}$ in Eq. (\ref{s8}) because this value is the
central energy density of the maximum mass obtained with the NJL model
for non-magnetized matter with the HK parameter set \cite{hatsuda}.
As in the MIT case, the magnetic field is taken as constant in $P_f$ and $P_l$ and
Eq.(\ref{s8}) is used in the terms proportional to $B^2$ only and the same holds for the
energy density terms.

\begin{figure}
\centering
\includegraphics[width=0.34\textwidth, angle =270 ]{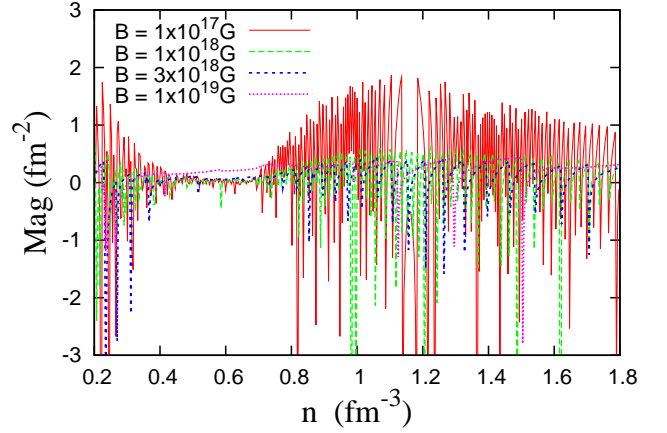}
\caption{Magnetization as a function of the number density  for the
  NJL model.}
\label{figmagnjl}     
\end{figure}

In Fig. \ref{figmagnjl} we show the magnetization of the system for
different values of the magnetic field. As already discussed in
\cite{ani} and also seen in Fig. \ref{figmitmag}, the number of van
alphen oscillations, related to the number of spikes in the NJL model,
are larger for lower magnetic fields and decrease for stronger
magnetic fields, when the number of filled Landau levels is small. 
Whenever  $\mu_f=s_f$, i.e., whenever $B$ approaches a $n\ne 0$ Landau level,
the denominator in Eq. (\ref{phimed}) becomes zero and the
contribution of this term to the magnetization of the system generates
the spikes. This kind of contribution is not present in the MIT model
and hence, the overall pattern is quite different in both models.
Another important difference refers to the lack of oscillations in
between 0.5 and 0.7 fm$^{-3}$  for all the fields within the NJL
model. As pointed out in \cite{ani}, this behavior is due to the $u$
and $d$ quark restoration of chiral symmetry and before the 
onset of the $s$ quark.  Again, this behavior is not seen in the
MIT model, where chiral symmetry effects are not present.
Finally, it is worth mentioning that the MIT bag model only produces
negative magnetization for very low magnetic fields while the NJL
model gives negative values for the magnetization for all fields. Bearing in
mind that negative magnetization refers to  magnetic repulsion
(diamagnetism) as opposed to positive magnetization that referes to
magnetic attraction (paramagentism), the MIT and the NJL models show different
physical pictures.  We address this point again in the next section.

\begin{figure}
\centering
\includegraphics[width=0.34\textwidth, angle =270 ]{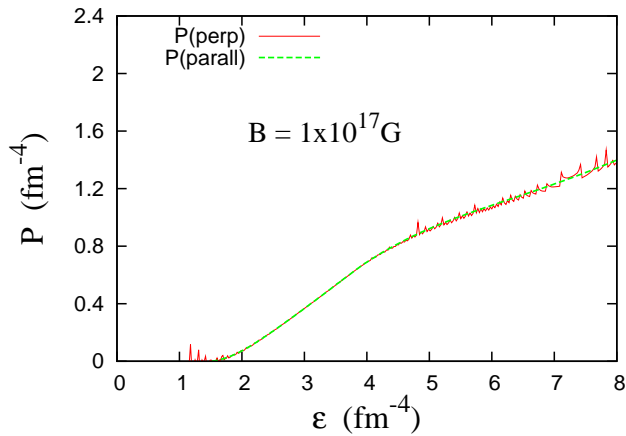}
\caption{NJL model EOS obtained for $B=10^{17}$ G without the term proportional to $B^2$}
\label{fignjlb17}   
\end{figure}

\begin{figure}
\centering
\includegraphics[width=0.34\textwidth, angle =270 ]{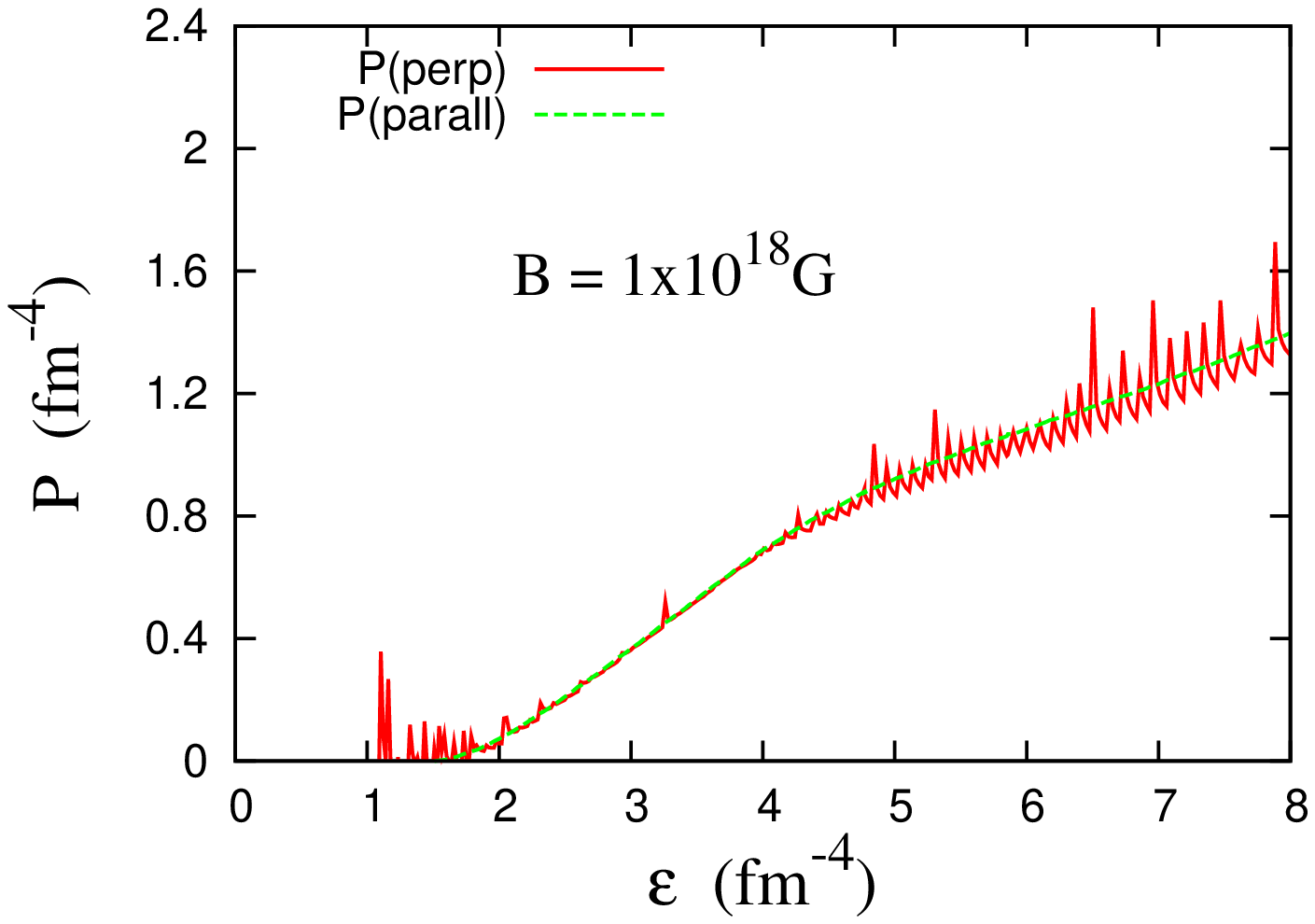}
\caption{NJL model EOS obtained for $B=10^{18}$ G without the term proportional to $B^2$}
\label{fignjlb18}   
\end{figure}

\begin{figure}
\centering
\includegraphics[width=0.34\textwidth, angle =270 ]{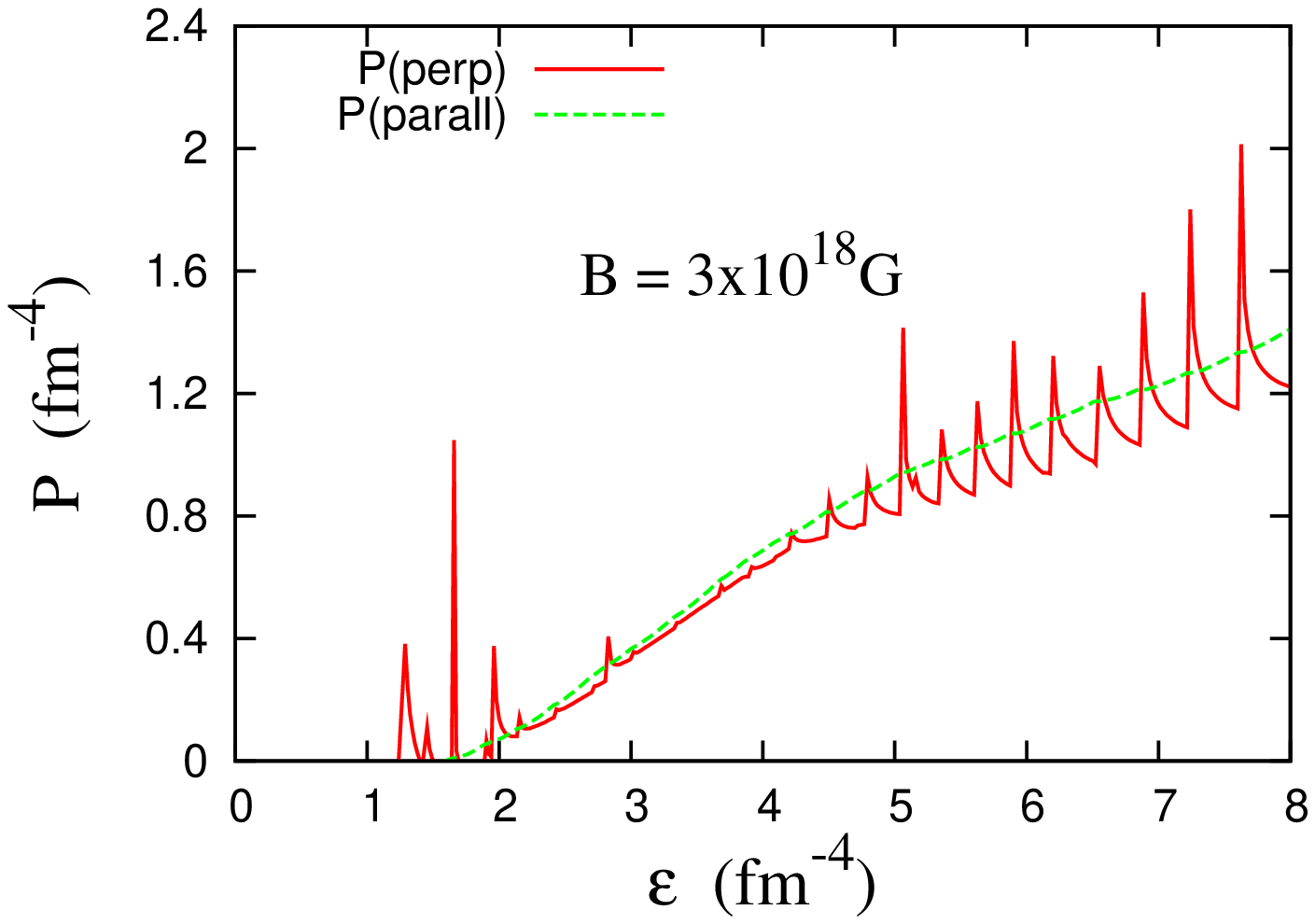}
\caption{NJL model EOS obtained for $B=3 \times 10^{18}$ G without the term proportional to $B^2$}
\label{fignjl3b18}   
\end{figure}

\begin{figure}
\centering
\includegraphics[width=0.34\textwidth, angle =270 ]{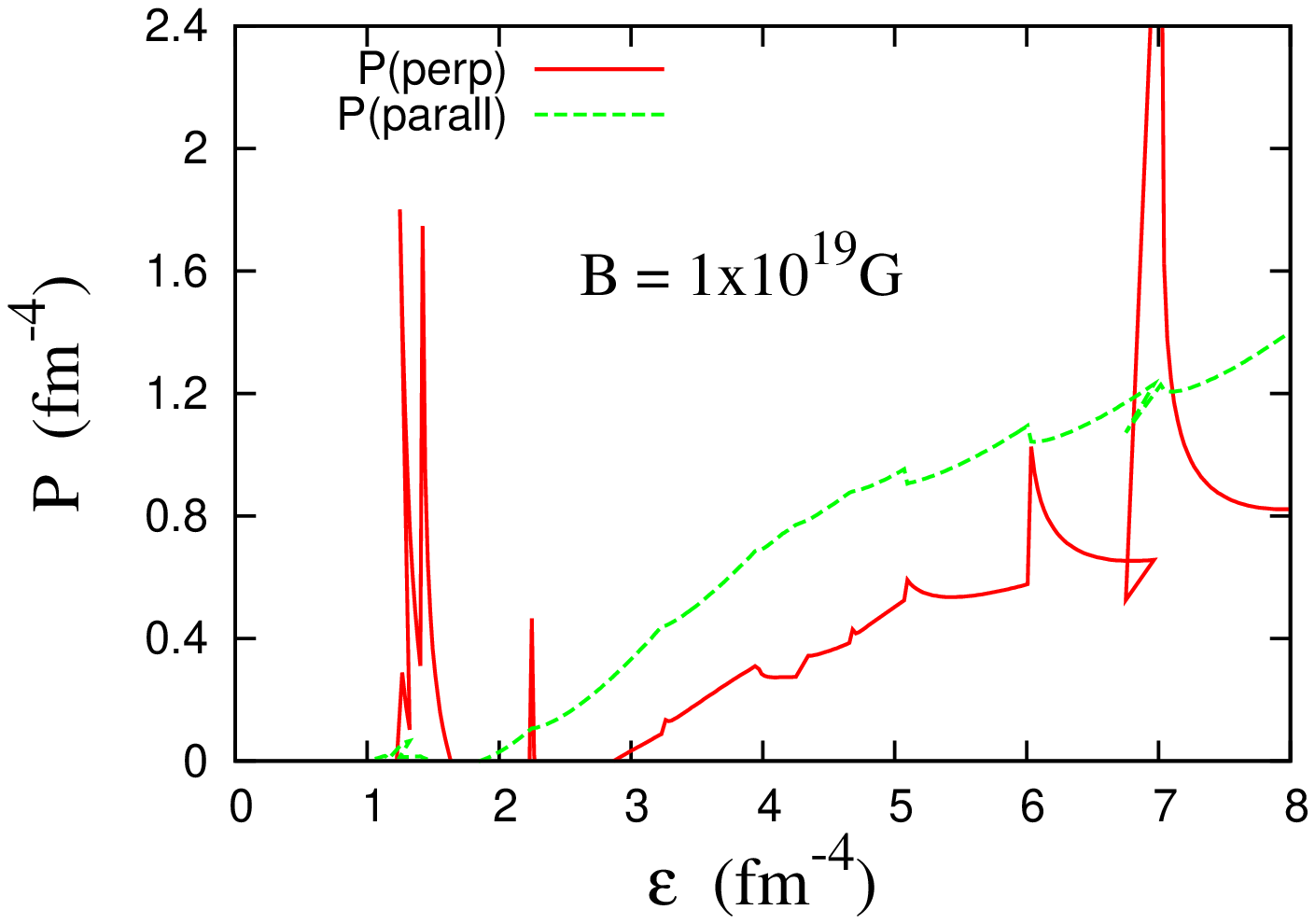}
\caption{NJL model EOS obtained for $B=10^{19}$ G without the term proportional to $B^2$}
\label{fignjlb19}   
\end{figure}

\begin{figure}
\centering
\includegraphics[width=0.34\textwidth, angle =270 ]{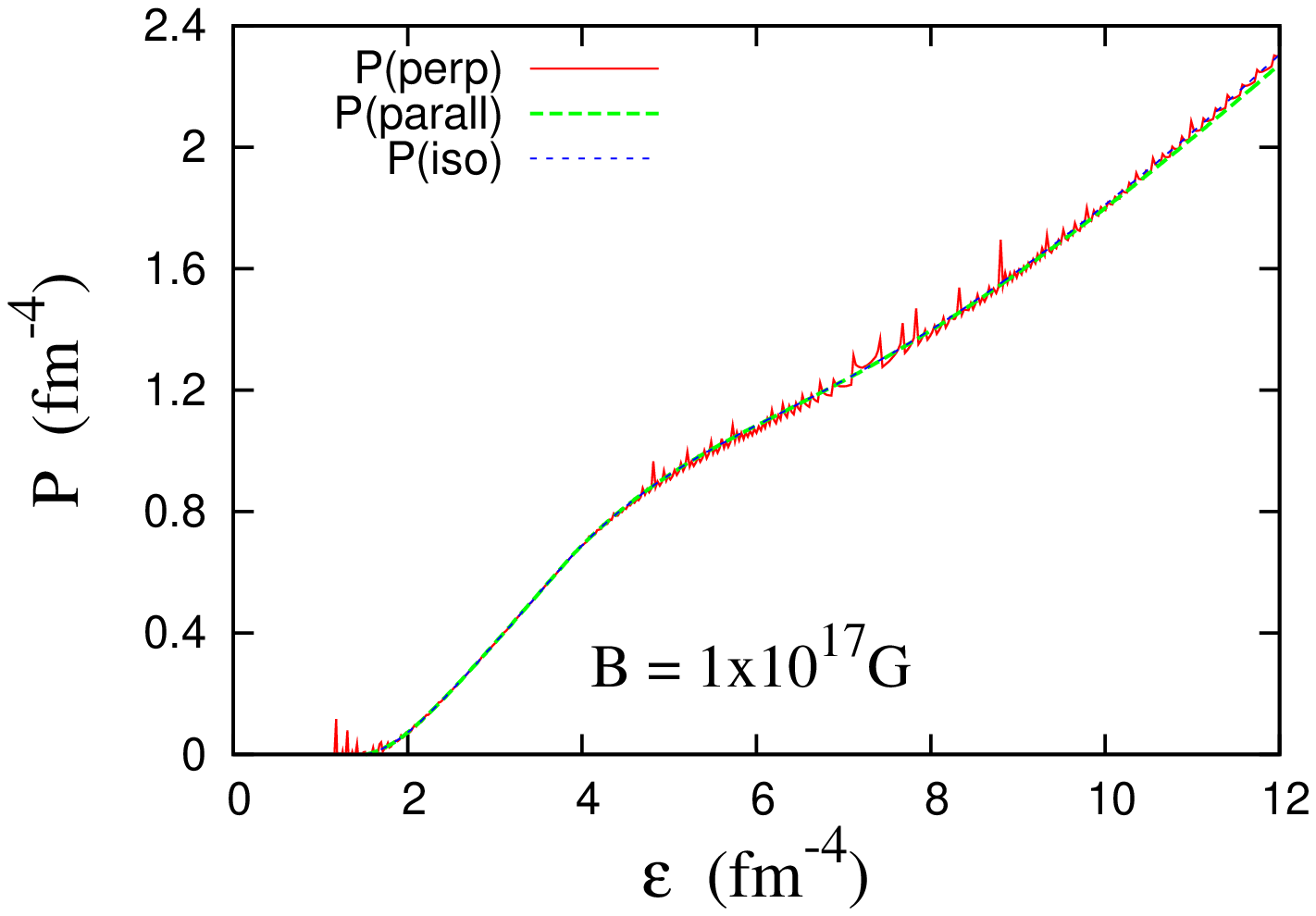}
\caption{NJL model EOS obtained for $B=10^{17}$ G with the term proportional to $B^2$}
\label{fignjlb17a}      
\end{figure}

\begin{figure}
\centering
\includegraphics[width=0.34\textwidth, angle =270 ]{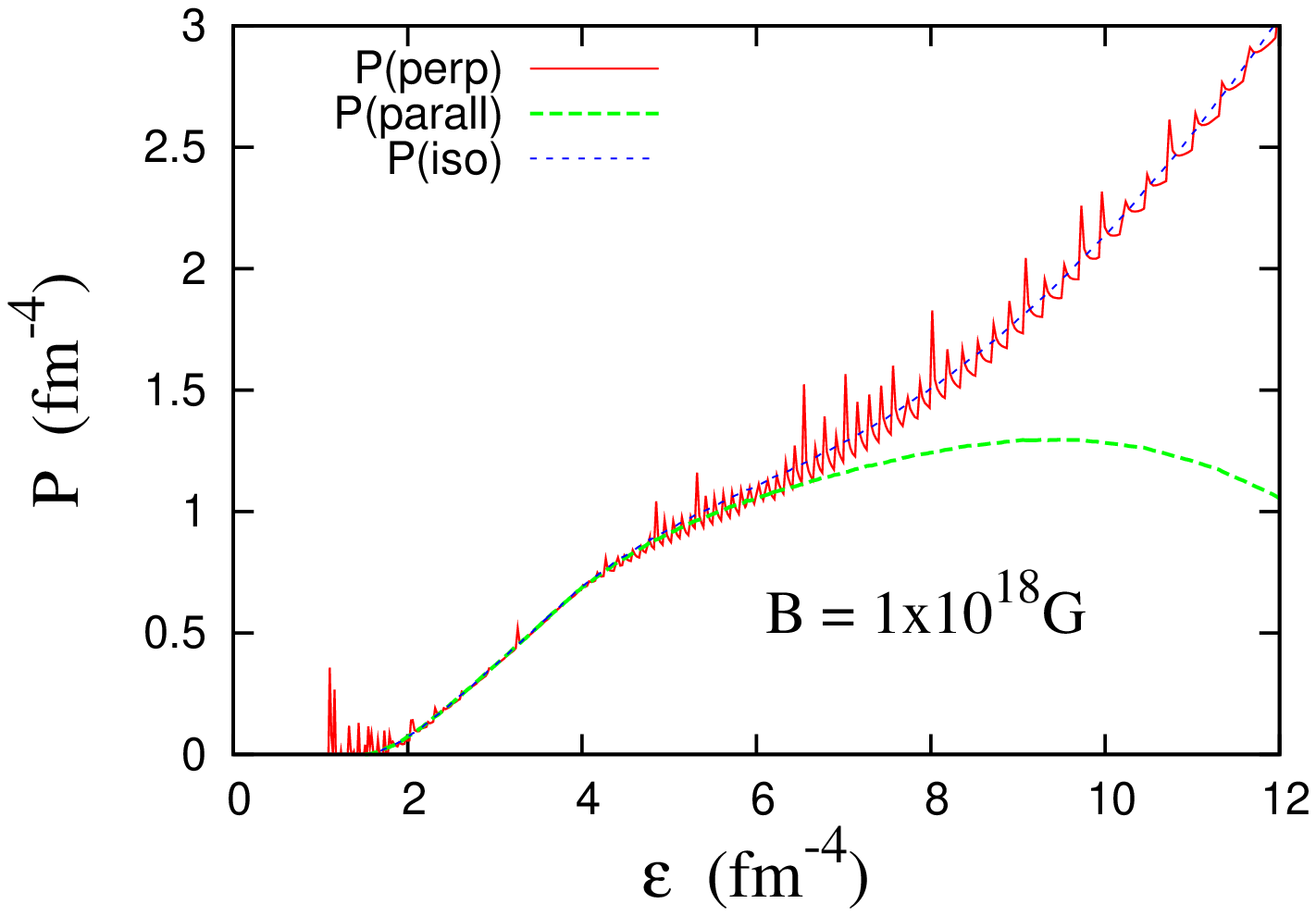}
\caption{NJL model EOS obtained for $B=10^{18}$ G with the term proportional to $B^2$}
\label{fignjlb18a}      
\end{figure}

\begin{figure}
\centering
\includegraphics[width=0.34\textwidth, angle =270 ]{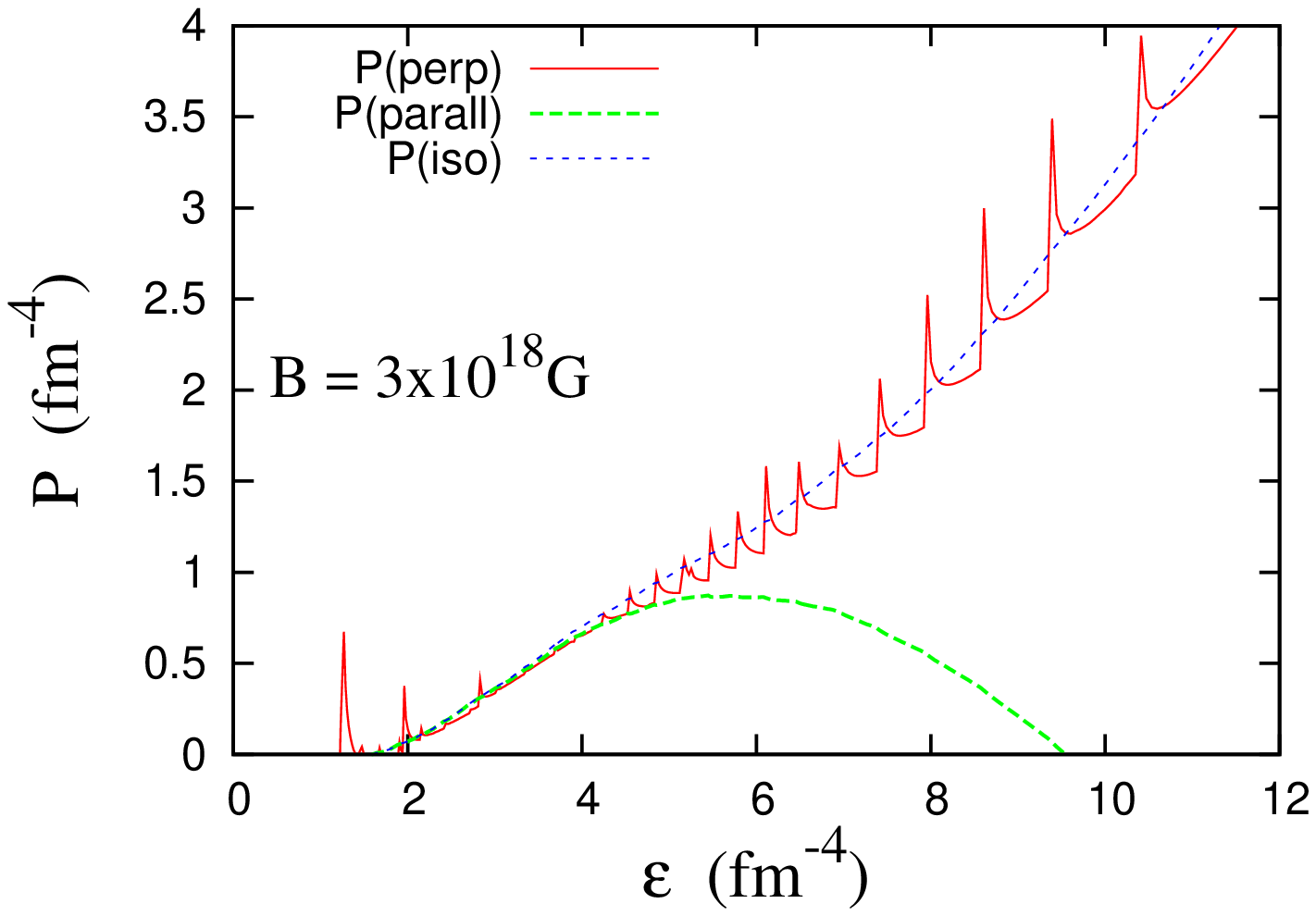}
\caption{NJL model EOS obtained for $B=3 \times 10^{18}$ G with the term proportional to $B^2$}
\label{fignjl3b18a}      
\end{figure}

\begin{figure}
\centering
\includegraphics[width=0.34\textwidth, angle =270 ]{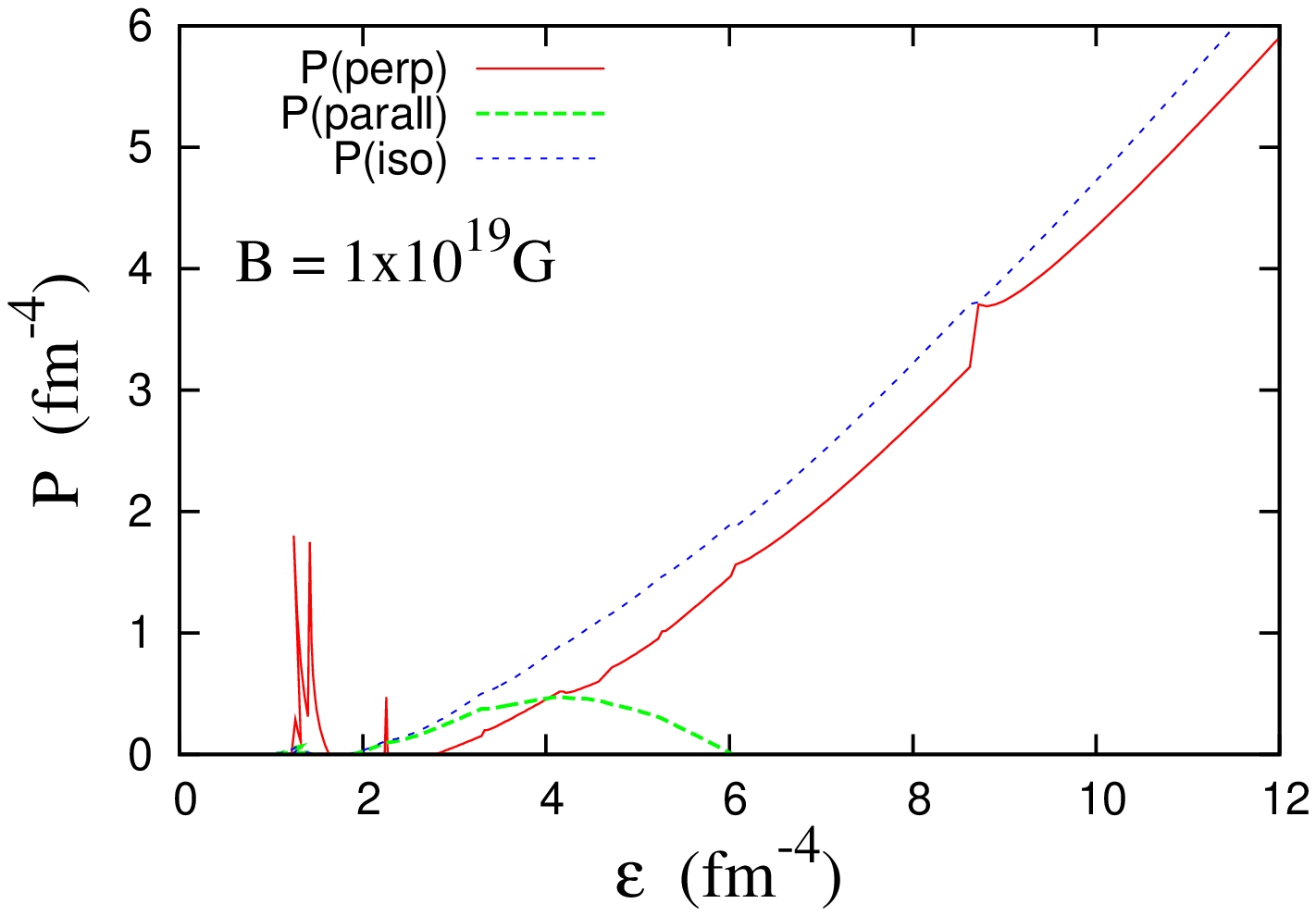}
\caption{NJL model EOS obtained for $B=10^{19}$ G with the term proportional to $B^2$}
\label{fignjlb19a}      
\end{figure}

In Figs. \ref{fignjlb17}, \ref{fignjlb18}, \ref{fignjl3b18} and
\ref{fignjlb19}, we show the anisotropic EOS with parallel and
prependicular pressures without the inclusion of the $B^2$ term and
in Figs. \ref{fignjlb17a}, \ref{fignjlb18a}, \ref{fignjl3b18a} and
\ref{fignjlb19a}, the same EOS are plotted with the inclusion of this
term. For $B=10^{17}$ G, the contribution of the $B^2$ term is almost
negligible and hence, we can see that the magnetization already plays
a role in the pressures, which are no longer coincident, as in the MIT
model. As $B$ increases, the effects of the magnetization are clearly
noticed. Once more, as in the MIT model, if the $B^2$ contribution is
taken into account, the parallel pressure
increases at lower energy densities and then decreases and goes to
zero, reaching this point at lower values for larger magnetic fields.
However, if one compares the isotropic pressure with the perpendicular
pressure, they are just similar in average, since the magnetization
effects are clearly present. This behavior is also quite different
from the one exhibited by the MIT model and discussed in a previous
section. Precisely because of the discontinuities in the perpendicular
pressure, the anisotropic EOS cannot be used as input to the TOV
equations.

{ At this point it is worth mentioning that similar calculations
  were perfomed in \cite{Huang_2015} for 
  the NJL model with and without a vector interaction. However, in \cite{Huang_2015} 
  the spikes in the EOS are not seem for fields as large as $4 \times
  10^{18}$ G .  We have checked that if we plotted the pressure and
  energy density in MeV/fm$^3$ (without the $B^2$ term) as in the
  mentioned reference, the spikes are almost imperceptible in the region of energy
densities corresponding to 500-1000 MeV/fm$^3$, but they
  are indeed present since they appear whenever $\mu_f=s_f$ in the
  denominator of Eq.(\ref{phimed}).  Notice
  also, that the parametrizaton of the electromagnetic interation in 
  the EOS (exactly the $B^2$ term) is different in both papers, resulting in different
  quantitative results. If this term grows too rapidly the
  oscillation-free zone can dominate the equation of state. }

\begin{figure}
\centering
\includegraphics[width=0.34\textwidth, angle =270 ]{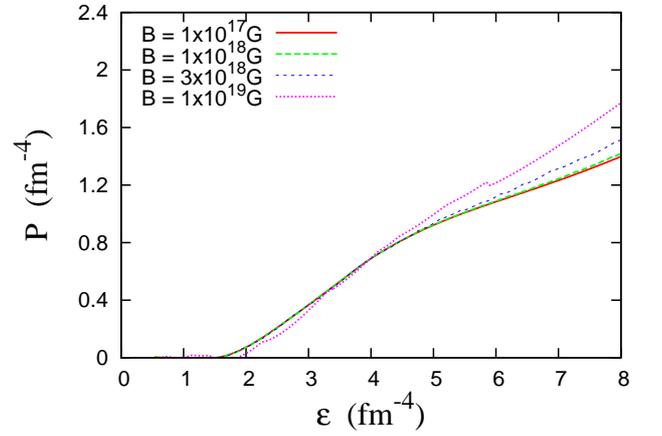}
\caption{NJL EOS for different values of the chaotic magnetic field.}
\label{fignjlchao}      
\end{figure}

\begin{figure}
\centering
\includegraphics[width=0.34\textwidth, angle =270 ]{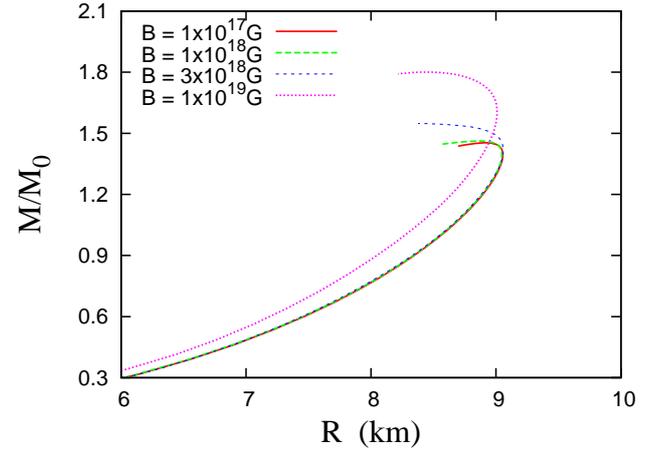}
\caption{NJL model: Mass-radius relation within isotropic pressure for different values of magnetic field.}
\label{tovisonjl}     
\end{figure}

\begin{figure}
\centering
\includegraphics[width=0.34\textwidth, angle =270 ]{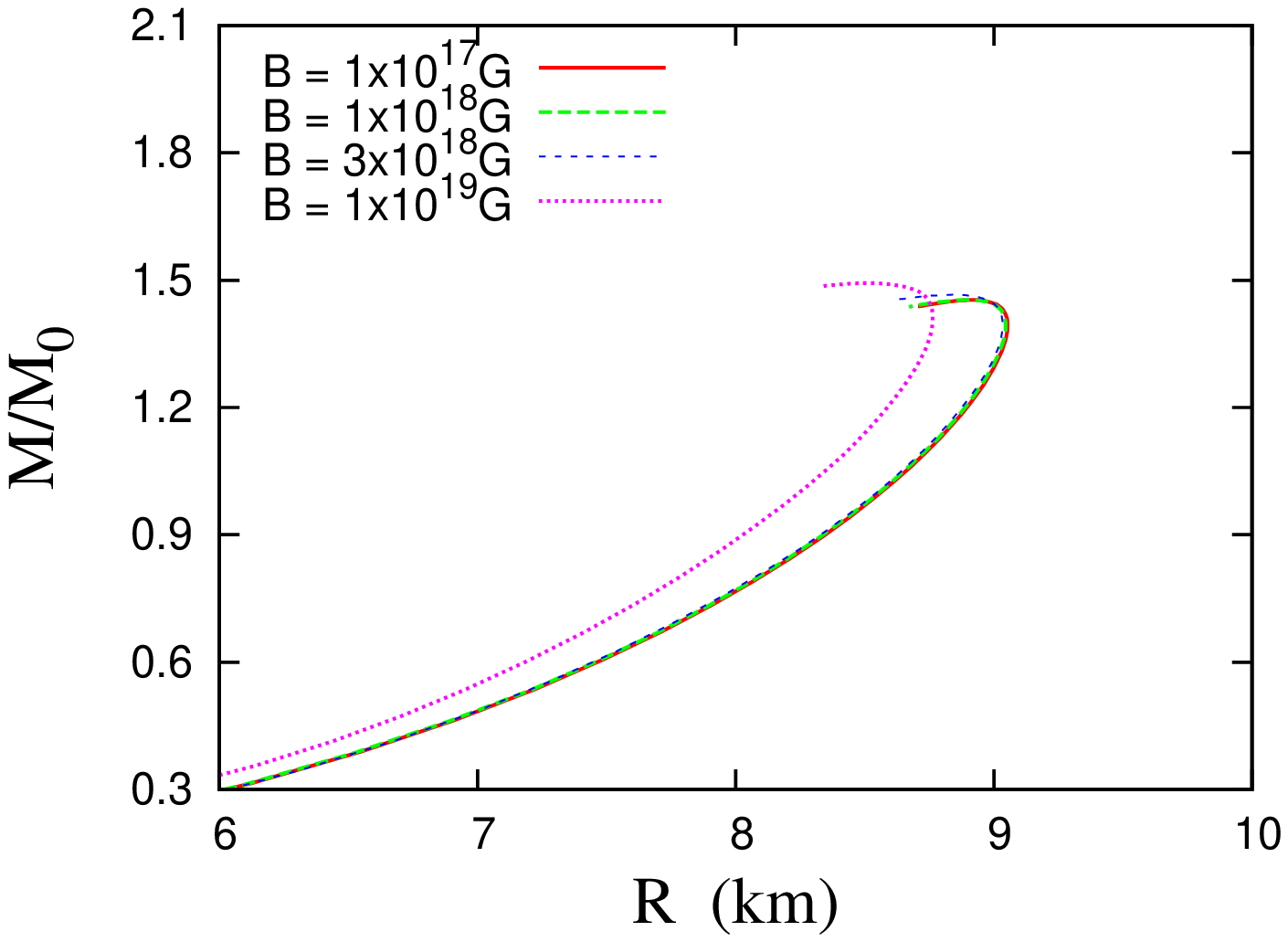}
\caption{NJL model: Mass-radius relation within chaotic magnetic field approximation for different values of magnetic field.}
\label{tovnjlcmf}     
\end{figure}

\begin{table}[t]
\caption{Properties of the maximum mass quark star}
\label{tabnjl}       
\begin{tabular}{llllll}
\hline\noalign{\smallskip}
Model & $B$ & $M_{max}$ & $Mb_{max}$ & R & ${\cal E}_c$ \\
\noalign{\smallskip}\hline\noalign{\smallskip}
& (G) & ($M_\odot$) & ($M_\odot$) & (km) & (fm$^{-4}$) \\
\noalign{\smallskip}\hline
NJL$_{\rm isotropic}$ & $10^{17}$ & 1.45  & 1.53  & 8.93 & 7.51 \\
\noalign{\smallskip}\hline
NJL$_{\rm chaotic}$ & $10^{17}$ & 1.45  & 1.53  & 8.92 & 7.66 \\
\noalign{\smallskip}\hline
\noalign{\smallskip}\hline
NJL$_{\rm isotropic}$ & $10^{18}$ & 1.46  & 1.54  & 8.87 & 7.76 \\
\noalign{\smallskip}\hline
NJL$_{\rm chaotic}$ & $10^{18}$ & 1.45  & 1.53  & 8.92 & 7.46 \\
\noalign{\smallskip}\hline
\noalign{\smallskip}\hline
NJL$_{\rm isotropic}$ & $3. 10^{18}$ & 1.55  & 1.62  & 8.44 & 9.58 \\
\noalign{\smallskip}\hline
NJL$_{\rm chaotic}$ & $3.  10^{18}$ & 1.47  & 1.54  & 8.85 & 7.86 \\
\noalign{\smallskip}\hline
\noalign{\smallskip}\hline
NJL$_{\rm isotropic}$ & $10^{19}$ & 1.80  & 1.77  & 8.42 & 10.25 \\
\noalign{\smallskip}\hline
NJL$_{\rm chaotic}$ & $10^{19}$ & 1.49  & 1.49  & 8.51 & 8.93 \\
\noalign{\smallskip}\hline
\end{tabular}
\vspace*{0.8cm}  
\end{table}

\begin{table}
\caption{Properties of the 1.4$M_{\odot}$ quark star}
\label{NJL14}       
\begin{tabular}{llllll}
\hline\noalign{\smallskip}
Model & $B$ & $M_{max}$ & $Mb_{max}$ & R & ${\cal E}_c$ \\
\noalign{\smallskip}\hline\noalign{\smallskip}
& (G) & ($M_\odot$) & ($M_\odot$) & (km) & (fm$^{-4}$) \\
\noalign{\smallskip}\hline
NJL$_{\rm isotropic}$ & $10^{17}$ & 1.40  & 1.46  & 9.05 & 4.82 \\
\noalign{\smallskip}\hline
NJL$_{\rm chaotic}$ & $10^{17}$ & 1.40  & 1.46  & 9.05 & 4.82 \\
\noalign{\smallskip}\hline
\noalign{\smallskip}\hline
NJL$_{\rm isotropic}$ & $10^{18}$ & 1.40  & 1.46  & 9.04 & 4.79 \\
\noalign{\smallskip}\hline
NJL$_{\rm chaotic}$ & $10^{18}$ & 1.40  & 1.46  & 9.05 & 4.82 \\
\noalign{\smallskip}\hline
\noalign{\smallskip}\hline
NJL$_{\rm isotropic}$ & $3. 10^{18}$ & 1.40  & 1.46  & 9.05 & 4.60 \\
\noalign{\smallskip}\hline
NJL$_{\rm chaotic}$ & $3.  10^{18}$ & 1.40  & 1.46  & 9.04 & 4.80 \\
\noalign{\smallskip}\hline
\noalign{\smallskip}\hline
NJL$_{\rm isotropic}$ & $10^{19}$ & 1.40  & 1.45  & 8.91 & 4.08 \\
\noalign{\smallskip}\hline
NJL$_{\rm chaotic}$ & $10^{19}$ & 1.40  & 1.43  & 8.77 & 5.01 \\
\noalign{\smallskip}\hline
\end{tabular}
\vspace*{0.8cm}  
\end{table}

Finally, in Fig. \ref{fignjlchao}, the EOS obtained with the chaotic
formalism is shown for different values of the magnetic field.
As the magnetic field increases the EOS becomes harder and within the
NJL model, they are more sensible to the magnetic field than within
the MIT model, where the EOSs just differ from each other for fields
of the order of $10^{19}$ G, as seen in Fig. \ref{figmitchao}.

We then use the EOS obtained with the isotropic and chaotic formalisms
to compute the stellar properties and the results are displayed
 in Figs \ref{tovisonjl}, and \ref{tovnjlcmf} respectively, and in Table \ref{tabnjl}
we summarize the properties of the maximum mass quark star.
As already expected, from the results existing in the literature
\cite{prc2}, \cite{prc1}, the maximum stellar masses are very low,
indicating that the NJL model cannot explain very massive
magnetars. A solution to this setback is to include the vector
interaction in the NJL model as in \cite{njlv}. 
The overall conclusions coming out of the macroscopic properties for
the NJL model are the same as the ones obtained with the MIT model,
i.e., the maximum mass increases much more from a
non-magnetized star to a strongly magnetized one if the isotropic
formalism is used than if the chaotic field approximation is assumed.

Finally we display the properties of the canonical 1.4$M_{\odot}$
neutron stars in Table \ref{NJL14}. We see that with the NJL, the quark star
radii are very low, and are always in accordance with ref.~\cite{webb},
instead the constraints of ref.~\cite{Steiner3}.

\section{MIT versus NJL - differences}
\label{sec:4}

\begin{figure}
\centering
\includegraphics[width=0.34\textwidth, angle=270 ]{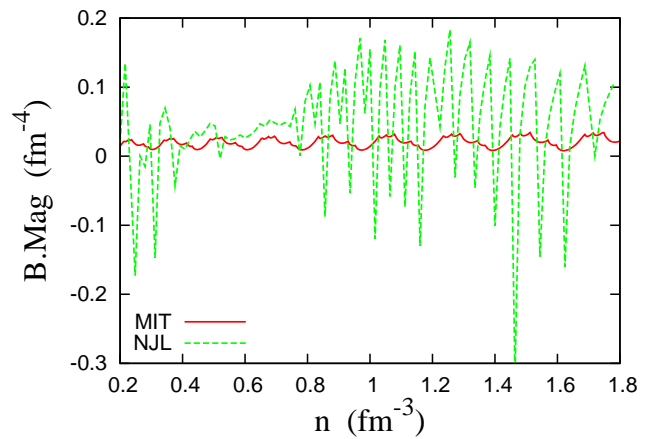}
\caption{Magnetic field times the system magnetization for $B=3 \times
  10^{18}$ G for the MIT and the NJL models.}
\label{figmagB}     
\end{figure}

 We start by reanalyzing the effects of the magnetization in the
perpendicular pressure of both models. According to
Eq. (\ref{pressmitaniso}), the magnetization always comes multiplied
by the magnetic field. We then plot, in Fig. \ref{figmagB} the term
that really influences the perpendicular pressure for $B=3 \times
10^{18}$ G. Theses results help us to understand the effects observed
in Figs. \ref{figmit3b18} and \ref{fignjl3b18} for the MIT and NJL models respectively.
The term proportional to the magnetization is very small in both
models and even when it becomes negative, the perpendicular pressure
is not much smaller than the parallel one.

\begin{figure}
\centering
\includegraphics[width=0.34\textwidth, angle =270 ]{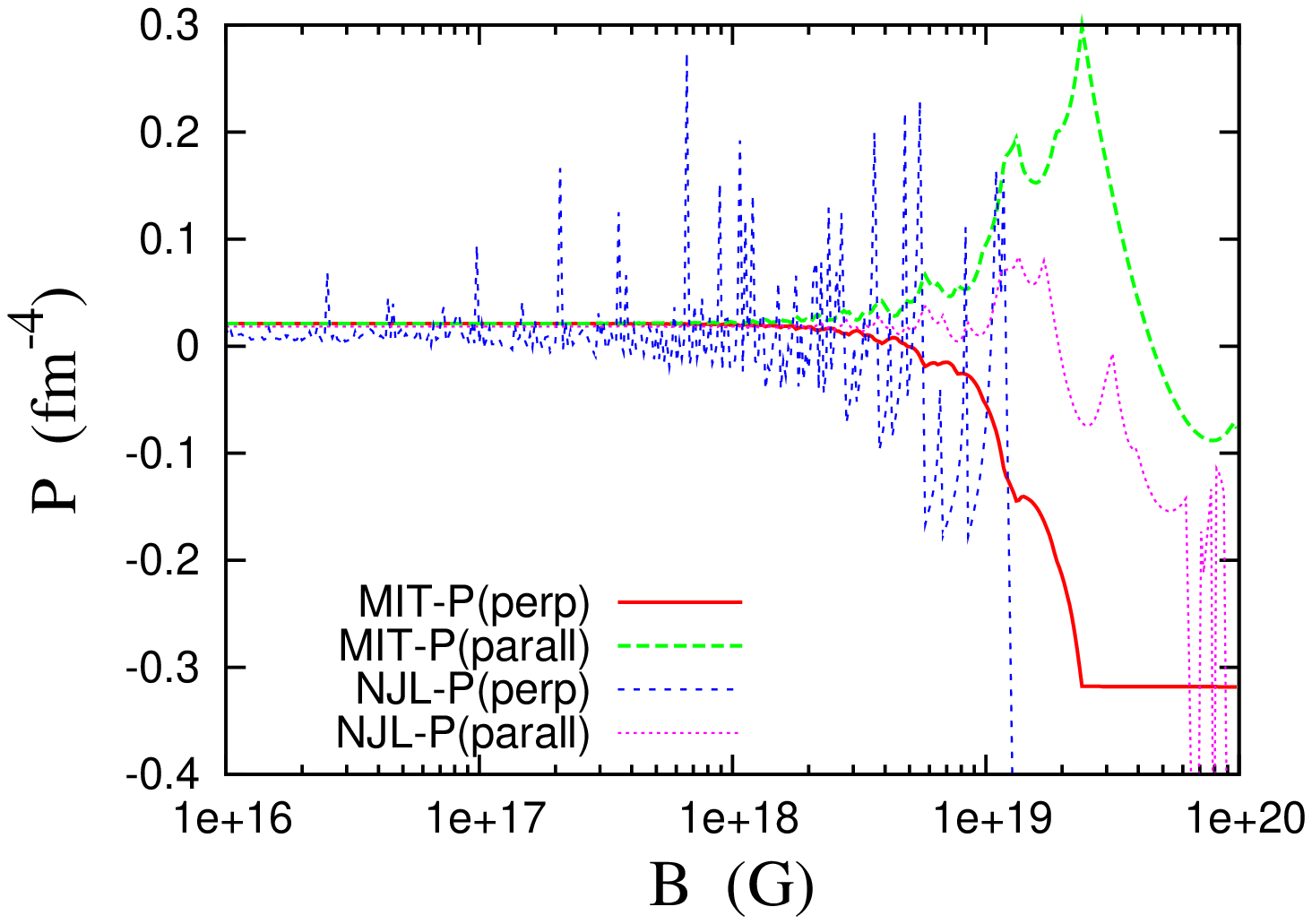}
\caption{Anisotropic pressures for the MIT and the NJL without the $B^2$ contribution for $n=0.2n_0$. }
\label{figB}     
\end{figure}

\begin{figure}
\centering
\includegraphics[width=0.34\textwidth,angle=270]{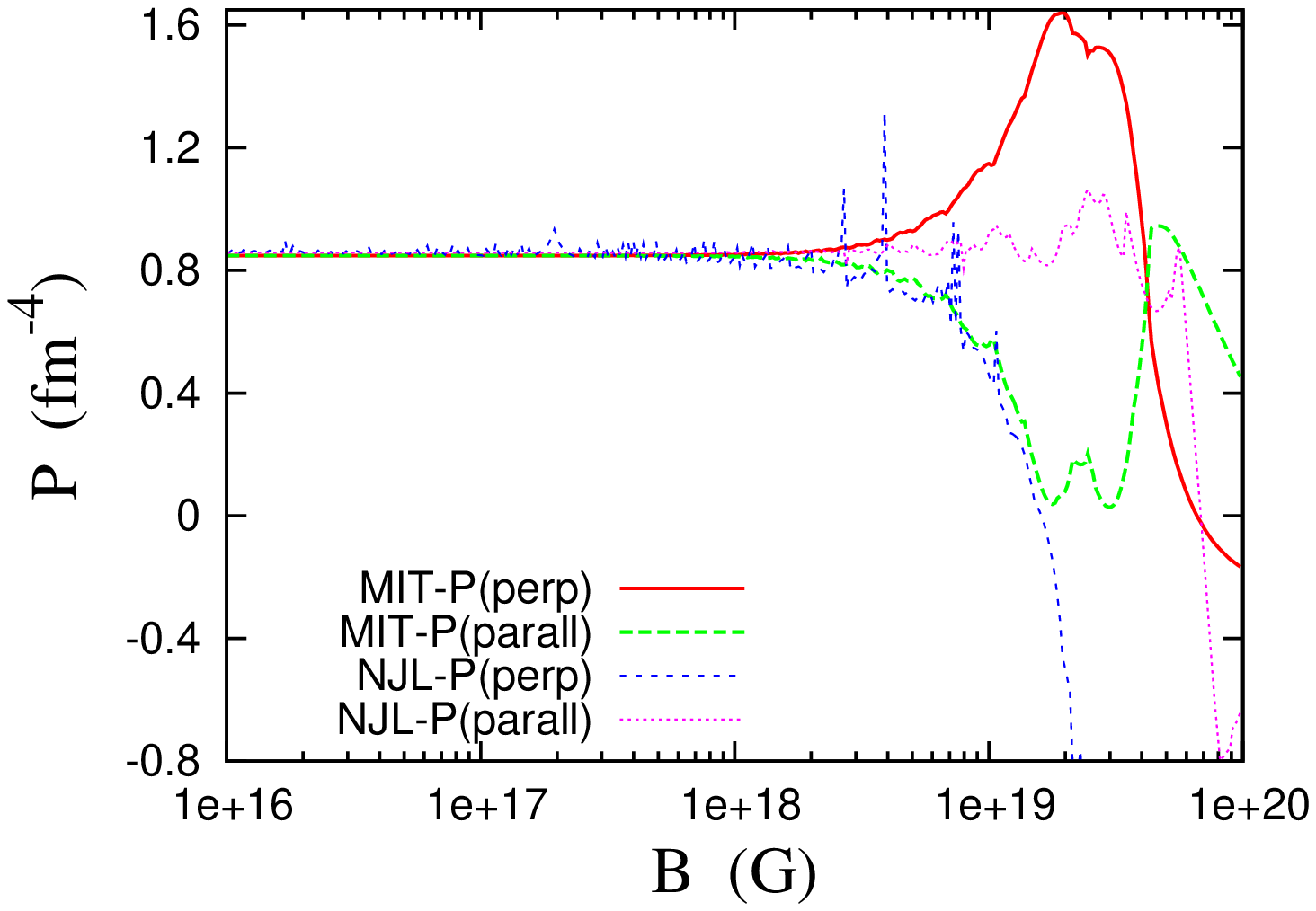}
\caption{Anisotropic pressures for the MIT and the NJL without the
  $B^2$ contribution for $n=5 n_0$. }
\label{figB2}     
\end{figure}

Our last figures, Fig. \ref{figB} and Fig. \ref{figB2}, show the
perpendicular and parallel pressures
obtained from both models at a fixed densities, $0.2 n_0$ and $5
n_0$ with $n_0=0.16$ fm$^{-3}$  
being the nuclear matter saturation density.
These densities are chosen to represent stellar matter close to the
crust and in the core of the star respectively.
In these graphs the terms proportional do $B^2$ are not included.
As shown in \cite{aurora1} for the MIT model, both pressures do not
differ much up to a magnetic field of the order of $10^{18}$ G, when
they start to deviate. For larger magnetic fields, of interest, for
instance, in heavy ion collisions, the deviation first increases and
then decreases again, stabilizing around $10^{20}$ G. For the NJL model,
the behavior is not very different in average. For the case of lower
density,  the perpendicular
pressure presents the typical spikes generated by the term that
includes the magnetization and the  parallel pressure is lower than
the corresponding parallel pressure obtained with the MIT model, but
the general trend is the same in both models. 
For $n= 5 n_0$, the number of spikes obtained with the NJL model
decrease because this density 
corresponds to the region where the magnetization is very low, as
shown in Fig. \ref{figmagnjl}. One can see that the larger the
density, the larger the differences between the pressures 
for fields higher than $10^{18}$G.

Another topic of interest is the effect of the magnetic field on the
quark star radii. We can also see from Tables 2 and 4, that 
the NJL model always yields lower maximum masses and 
lower radii for the canonical 1.4$M_{\odot}$ NS. Comparing Figs. \ref{tovmitiso} and \ref{tovmitcmf} with
Figs. \ref{tovisonjl} and \ref{tovnjlcmf} we see that while the magnetic field
increases the maximum mass in both MIT and NJL models, the effect on the radii is opposed.
In both formalisms, isotropic and with chaotic field, for a fixed
mass, the radii increase with the increase of the magnetic field in
the MIT bag model and decrease in the NJL. In fact, the only case that produces a more compact
quark star in the MIT bag model is when anisotropic pressure is
considered. Unfortunately we are unable to calculate the TOV equation
with the anisotopic NJL EOS 
due to the high oscillations in the pressure. Nevertheless, even if we could,
these results would have to be taken with care since, as discussed in
Section \ref{sec:2} because just one of the pressures is taken into
account, since the other one becomes zero at still quite low densities.
To trust macroscopic results
obtained from a TOV-like equation, a self-consistent model for the
study of the quark star structure such as the one proposed in
\cite{Micaela} should be done, so that more precise conclusions could
be drawn.

\section{Final remarks}

In the present work we have used two models widely used to describe
quark matter, the MIT and the NJL model and checked the effects of
strong magnetic fields on their magnetization, EOS and stellar
properties. To accomplish this task, we have revisited three formalisms
generally applied to describe magnetized matter: the first one assumes
that the EOS is isotropic, the second one that it is anisotropic and
the third one that a chaotic field is possibly generated and
maintained in the system.

 For the second formalism, the calculation of
the magnetization of the system is mandatory, since one of the
pressures depends on this quantity. However, independently of the
formalism used, the magnetization is a quantity that should be
understood in a magnetized system. We have verified that both models
produce quite different results. The NJL model presents three terms
for the magnetization coming from the medium, the vacuum
and the magnetic field itself. Just the first one is present in the
MIT model and this explains most of the differences between both
models.

As for the EOS, the isotropic and the chaotic formalisms present
similar patterns in both models, the NJL being always slightly more
sensible to the magnetic field strength. However, the anisotropic EOS  
shows a quite discontinuous behavior due to the spikes that appear in
the calculation of the magnetization of the system. 

When the macroscopic properties are computed, the isotropic EOSs always
provide maximum stellar masses that are much larger than the ones
obtained with a non-magnetized matter, in contrast with the values
obtained with the chaotic field that do not result in much higher
maximum masses. The small increase in the maximum masses is also found
in more sophisticated calculations \cite{Mallick2}, \cite{Micaela}.

Other important aspects related to quark matter subject to strong
magnetic fields are the polarization and the viscosity of the system. 
For totally polarized matter, all particles lie on the lowest Landau
level. The quark spin polarization has already been studied for both the MIT
bag model in \cite{aurora1} and the NJL model in \cite{njlt},
\cite{ani}  and hence, we only comment on the already known results.
The polarization of the system increases with the increase of the
magnetic field. In \cite{aurora1}, it is shown that the total
polarization obtained with the MIT bag model occurs for $B \simeq 2
\times 10^{19}$ G for matter in $\beta$-equilibrium. This result
depends on the density and the temperature considered  and is
different for each quark flavor, since it depends on the charge of the
quark \cite{njlt}. For matter in $\beta$-equilibrium described by the
NJL model, the electron becomes totally polarized for magnetic fields
larger than $9 \times 10^{17}$ G and the $s$ quark before $10^{18}$ G.
The $u$ and $d$ quarks become totally polarized for approximately the
same magnetic field as in the MIT bag model, i.e., higher than
$10^{19}$ G and lower than $10^{20}$ G, the exact value being
parameter dependent. 

For quark matter, the non-leptonic processes 
$$ u + s \leftarrow u + d, \quad u + d \leftarrow u + s$$
are responsible for the restoration of chemical equilibrium and hence,  
they determine the bulk viscosity of the system. In \cite{sedrakian},
bulk (and shear) viscosities are derived and discussed. Similar
calculations remain to be done with the MIT bag model for the chaotic
field approximation and for the NJL model and the results can shed
some light on the stability of magnetized quark matter. 

The physics of quark star radii is another open puzzle. It is well known that
the radii of hadronic neutron stars are correlated with the symmetry energy
slope $L$, \cite{horowitz}, \cite{rafael}, \cite{Gandolfi},\cite{Constança},\cite{Lopes3}. However
we do not know yet if 
there is an analogue physical parameter that controls the quark star radius.

As mentioned in the Introduction, the influence of the quark AMM
should not be completely disregarded. In \cite{chang2011}, 
it was shown that for quark matter the scale
for the perturbative approach is defined by the constituent quark mass
and therefore, the effects of the  AMM should
also be considered in the description of magnetized quark matter.
As the quark mass is about one third the nucleon mass,
the critical field will be approximately one order of magnitude
smaller than the nucleon critical field, but still larger than the maximum
field expected inside a quark star. Hence, the degeneracy of Landau
levels will certainly be affected by the inclusion of the coupling between the field and the fermion
anomalous magnetic moment. In a future work this problem should also
be considered.

Before we conclude, it is important to stress that we have not
exhausted the discussion on all formalisms since, as already mentioned
in the Introduction, there are still other ones in the literature. 
One possible approach to deal with the anisotropic pressures and avoid
the use of the TOV equations,  is to follow the prescription used in
\cite{Mallick2} with the help of the Hartle-Thorn method.
We have seen that effects of the magnetization should not be disregarded 
as far as magnetic fields are stronger than $10^{17}$ G and hence, the
calculations performed in \cite{Mallick2}, which completely ignore the
contribution due to the magnetization of the system in the
perpendicular pressure, should be revisited.
There are also more sophisticated calculations that consider the anisotropy
in solving  Einstein's field equations in a fully general relativistic
formalism~\cite{Bocquet},\cite{Cardall},\cite{Micaela}, but the
computational price paid is huge. Moreover, the fields generated at
the surface of the star in \cite{Micaela} are of the order of
$10^{17}$ G, much higher than the $10^{15}$ G magnetic field
expected in the surface of magnetars.

{\bf Acknowledgements} D.P.M. acknowledges fruitful discussions with
Marcus Benghi Pinto, Constan\c ca Provid\^encia and Ver\^onica
Dexheimer Strickland.  This work was partially supported by
Conselho Nacional de Desenvolvimento Cient\'{\i}fico e Tecnol\'{o}gico (CNPq-Brazil),
CEFET/MG and by Funda\c c\~{a}o de Amparo \`{a} Pesquisa e Inova\c c\~{a}o do
Estado de Santa Catarina (FAPESC-Brazil), 
under project 2716/2012.

\end{document}